\documentclass[aps,11pt,nofootinbib,preprintnumbers,groupedaddress,superscriptaddress]{revtex4-2}
\usepackage{amssymb, amsfonts, amsmath, bm, empheq}
\usepackage[%dvipdfmx
]{graphicx}
\usepackage{color}
\usepackage{mathrsfs}
\usepackage{enumerate}
\usepackage{multirow}
\usepackage[%dvipdfmx
]{hyperref}
\hypersetup{% 
 setpagesize=false,
 bookmarksnumbered=true,%
 bookmarksopen=true,%
 colorlinks=true,%
 linkcolor=blue,%
 citecolor=red}
\begin{document}
\title{
Periapsis shifts in dark matter distribution around a black hole
}
\author{Takahisa Igata}
\email{takahisa.igata@gakushuin.ac.jp}
\affiliation{
Department of Physics, Gakushuin University,
Mejiro, Toshima, Tokyo 171-8588, Japan}
\affiliation{KEK Theory Center, 
Institute of Particle and Nuclear Studies, 
High Energy Accelerator Research Organization, Tsukuba 305-0801, Japan}
\author{Tomohiro Harada}
\email{harada@rikkyo.ac.jp}
\affiliation{Department of Physics, Rikkyo University, Toshima, Tokyo 171-8501, Japan}
\author{Hiromi Saida}
\email{saida@daido-it.ac.jp}
\affiliation{Daido University, Nagoya, Aichi 457-8530, Japan}
\author{Yohsuke Takamori}
\email{takamori@wakayama-nct.ac.jp}
\affiliation{National Institute of Technology (KOSEN), 
Wakayama College, Gobo, Wakayama 644-0023, Japan}
\date{\today}
\preprint{KEK-Cosmo-0283, KEK-TH-2381, RUP-22-3}

\begin{abstract}
We consider the periapsis shifts of bound orbits of stars
on static clouds around a black hole. The background spacetime is constructed 
from a Schwarzschild black hole surrounded by a static and spherically symmetric self-gravitating system of massive particles, 
which satisfies all the standard energy conditions and physically models the gravitational effect of dark matter 
distribution around a nonrotating black hole.
Using nearly circular bound orbits of stars, we 
obtain a simple formula for the precession rate.
This formula explicitly shows that the precession rate is determined by 
a positive contribution (i.e., a prograde shift) from the conventional general-relativistic effect and a negative contribution (i.e., a retrograde shift) from the local matter density.
The four quantities
for such an orbit (i.e., the orbital shift angle, the radial oscillation period, the redshift, and the star
position mapped onto the celestial sphere) determine the local values of the background model functions.
Furthermore, we not only evaluate the precession rate of nearly circular bound
orbits in several specific models 
but also
simulate several bound
orbits with 
large eccentricity and their 
periapsis shifts. 
The present exact model demonstrates that the retrograde precession does not mean any exotic central objects such as naked singularities or wormholes but simply 
the existence of significant energy density of matters on the star orbit around the black hole. 
\end{abstract}
\maketitle

%%%%%%%%%%
\section{Introduction}
\label{sec:1}
%%%%%%%%%%
We are currently in the midst of rapid progress in observing the vicinity of black holes.
In particular, in the observation of the center of our galaxy, 
the orbital evolution of the so-called S-stars orbiting a supermassive black hole candidate, Sagittarius A$^\ast$ (Sgr A$^\ast$), has been actively investigated~\cite{Ghez:2000ay,Schodel:2002py}. 
Because S-stars can be regarded as test particles on the gravitational field of Sgr A$^\ast$,
precise measurements of their orbital evolution provide information on the central object and its surrounding matters, as well as on the spacetime geometry~\cite{Do:2019txf,Saida:2019mcz,Abuter:2020dou}.

A theoretical understanding of such orbital evolution is essential for identifying observables and interpreting observational results. 
One of the typical examples is that the elliptic orbit, well-known in Newtonian gravity, rotates in the same direction as the orbital evolution because of the 
general-relativistic effect~(see, e.g., Ref.~\cite{Weinberg:1972}). 
This is the so-called 
periapsis shift phenomenon of the 
bound orbits. 
Thus, by observing the displacement of the orbit due to this precessional motion, we can estimate the general-relativistic correction to the gravitational field. 
Furthermore, it has been discussed that supermassive and intermediate-mass black holes may be associated with large dark matter overdensities, called density spikes of dark matter~\cite{Gondolo:1999ef, Bertone:2009kj, Sadeghian:2013laa, Lacroix:2018zmg} and an ultralight dark matter solitons~\cite{Bar:2019pnz}.
Obviously, the contribution from
extended mass distribution
must also be considered as a correction to the black hole gravitational field. 
However, although it has been discussed in post-Newtonian gravity~\cite{Rubilar:2001} and by the general-relativistic Plummer model~\cite{Igata:2022nkt}, the effect of matter distribution on the periapsis shift phenomena is still controversial in the framework including a 
general-relativistic black hole.
Therefore, the most important next 
issue is to clarify the competition between the 
general-relativistic and 
local-density effects on the 
periapsis shift in a spacetime where a black hole and matter distribution are 
coexistent.
Such knowledge will broaden our understanding of the effects of matter fields on particle dynamics in the strong gravity regime. Besides, it will be useful for comparison with the case where the center is not a black hole but an exotic object~\cite{Bini:2005dy,Bambhaniya:2021ybs,Ota:2021mub}.

To clarify the above issues, we construct a background spacetime with a static and spherically symmetric distribution of massive particles
around the Schwarzschild black hole by exactly solving the Einstein equations.
Applying a method developed by Einstein~\cite{Einstein:1939, Geralico:2012jt}, 
we obtain a black hole spacetime surrounded by matter distribution known as an Einstein cluster, which 
consists of an averaged distribution of collisionless particles. 
Then, utilizing this black hole spacetime, we aim to formulate the competition between 
the general-relativistic and 
local-density effects that determine the direction of the 
periapsis shift
of the bound
geodesic orbits of stars on the matter distribution.
Using this formulation,
we consider the case in general relativity where the retrograde shift due to extended matter distribution can compensate for the prograde shift due to the
general-relativistic effect (see Refs.~\cite{Rubilar:2001,Nucita:2007qp, Zakharov:2007fj,
Iwata:2016ivt} for the post-Newtonian regime). 
However, it is not obvious whether extended matter distribution 
contributes to the retrograde shift in any case.
Therefore, it is quite important to discuss the 
periapsis shift
by considering the standard energy conditions and other physically reasonable conditions for the matter field. 

Here, we review recent progress in black hole spacetimes 
involving matter distribution constructed by what we call the Einstein cluster approach, 
a method for obtaining solutions to the Einstein equations in which the matter distribution has 
a stress-energy tensor of the same type as the Einstein clusters. 
As Boehmer and Harko~\cite{Boehmer:2007az} and Lake~\cite{Lake:2006pp} showed, 
an Einstein cluster was a possible galactic dark matter halo model. 
Using the Einstein cluster approach, 
Cardoso \textit{et al.} recently proposed a self-consistent black hole model immersed 
in a whole galactic halo with a Hernquist-type profile~\cite{Cardoso:2021wlq}.  
The effects of the halo on 
the innermost stable circular orbit, black hole shadow, and gravitational quasi-normal modes~\cite{Cardoso:2021wlq}; 
electromagnetic quasi-normal modes~\cite{Konoplya:2021ube}; 
and the epicyclic resonance of quasi-periodic oscillations in active galactic nuclei~\cite{Stuchlik:2021gwg} were discussed. 
Later, using a matter field subjected to a different equation of states, 
Jusufi proposed other exact solutions and considered the effect of black hole shadows and the stability of particle orbits~\cite{Jusufi:2022jxu}. 
More recently, Figueiredo \textit{et al.} developed a geometry with two density profiles to compare 
with the Hernquist-type profile, revealing variations of geodesic motion and gravitational wave fluxes 
and these model-independent nature~\cite{Figueiredo:2023gas}.

We should clarify the characteristics of our model compared to the previous models. 
The idea of using the Einstein cluster approach to surround a black hole with a matter field is common to these previous studies. 
Note, however, that the model in Ref.~\cite{Cardoso:2021wlq} 
has a serious problem in its physical interpretation because it
violates the dominant energy condition near the black hole, 
whereas, in the current paper, we improve the model so that it can
satisfy all of the dominant, weak, null, and strong energy conditions
over the entire region.

This paper is organized as follows. 
In Sec.~\ref{sec:2}, we review a static and spherically symmetric cloud solution to the Einstein equations. In particular, we construct 
a black hole surrounded by a
static Einstein cluster and clarify 
physically reasonable constraints based on the standard energy conditions.
In Sec.~\ref{sec:3}, we formulate 
the dynamics of a freely falling particle
in the Einstein cluster around a black hole and show the conditions for the existence of nearly circular 
bound orbits.
Then, we derive a formula that shows how the conflicting general-relativistic and 
local-density effects determine the precession rate that characterizes 
the periapsis shift.
In Sec.~\ref{sec:4}, we evaluate the precession rate 
for nearly circular orbits in 
several 
spacetime models obtained by giving concrete forms to the metric functions.
Furthermore, we demonstrate 
periapsis shifts of bound
orbits with 
large eccentricity.
In Sec.~\ref{sec:5}, we discuss how to determine the matter distribution around a black hole 
using the formula, given
observational data on the orbiting stars.
Section~\ref{sec:6} is devoted to 
a summary and discussion. 
We use units in which $G=1$ and $c=1$.

%%%%%%%%%%
\section{Static clouds around the Schwarzschild black hole}
\label{sec:2}
%%%%%%%%%%
We review a static and spherically symmetric cloud solution to the Einstein equations. 
Let $t$ be a static time, and $r$ be the areal radius.
The labels $(\theta, \varphi)$ are the standard spherical coordinates. 
Using these coordinates $x^\mu=(t, r, \theta, \varphi)$, we consider 
the general metric ansatz of static and spherically symmetric spacetimes,
\begin{align}
\label{eq:met2}
g_{\mu\nu}\:\!\mathrm{d}x^\mu\:\!\mathrm{d}x^\nu
=-\left(1-\frac{2\alpha(r)}{r}\right)\mathrm{d}t^2+\left(1-\frac{2m(r)}{r}\right)^{-1}\mathrm{d}r^2+r^2 (\mathrm{d}\theta^2+\sin^2\theta\:\!\mathrm{d}\varphi^2),
\end{align}
where $\alpha(r)$ and $m(r)$ are continuous functions of $r$, and in particular $m(r)$ is the Misner-Sharp mass~\cite{Misner:1964je,Hayward:1994bu}. 
Now we assume that 
\begin{align}
\label{eq:maspt}
&0\le m<\frac{r}{2},
\\
\label{eq:alppt}
&\alpha <\frac{r}{2}.
\end{align}

Before assuming the Einstein cluster of collisionless particles, we review the Einstein cluster approach, 
a method for obtaining solutions to the Einstein equations in which the matter distribution 
has a stress-energy tensor of the same type as the Einstein clusters. 
We consider the following form of the stress-energy tensor:
\begin{align}
\label{eq:Tmunu}
T^{\mu}{}_{\nu}=\mathrm{diag} (-\epsilon, 0, \Pi, \Pi),
\end{align}
where $\epsilon$ and $\Pi$ denote energy density and tangential pressure, respectively, and we assume that the radial pressure vanishes. 
Through the Einstein equations, $\epsilon$ and $\Pi$ are related to $m$ as 
\begin{align}
\label{eq:ep1}
\epsilon(r)&=\frac{m'}{4\pi r^2},
\\
\label{eq:PI1}
\Pi(r)&=\frac{m}{2(r-2m)} \epsilon(r),
\end{align}
and the vanishing radial pressure leads to the remaining nontrivial equation 
\begin{align}
\label{eq:alp1/2}
\alpha'=\frac{1}{2}\left(
1-\frac{r-2\alpha}{r-2m}
\right)
=\frac{\alpha-m}{r-2m},
\end{align}
or equivalently,
\begin{align}
\label{eq:malpha}
m=\frac{\alpha-\alpha' r}{1-2\:\!\alpha'},
\end{align}
where the prime denotes differentiation with respect to $r$.
As $m$ and $\alpha$ are continuous, the function $\alpha'$ is also continuous. 
Note that Eq.~\eqref{eq:alp1/2} together with the inequalities~\eqref{eq:maspt} and \eqref{eq:alppt} implies that 
\begin{align}
\label{eq:haltalp}
\alpha'<\frac{1}{2}. 
\end{align}
Consequently, for given $m(r)$ and $\alpha(r)$ 
under the constraints~\eqref{eq:maspt} and \eqref{eq:alppt},
we can specify the matter distribution $\epsilon(r)$ and $\Pi(r)$ through Eqs.~\eqref{eq:ep1} and \eqref{eq:PI1}, respectively. 
This static configuration is possible by a balance between the gravitational force and the tangential pressure.

For a matter field to be physically reasonable, the stress-energy tensor must satisfy several energy conditions (see, e.g., Ref.~\cite{Wald:1984}).
Imposing the energy conditions further restricts 
$m$ and $\alpha$ than Eqs.~\eqref{eq:maspt} and \eqref{eq:alppt}.
Some of the energy conditions for $T^\mu{}_\nu$
are written as follows: (\hspace{.18em}i\hspace{.18em}) weak energy condition, $\epsilon\ge 0$ and $\epsilon+\Pi \geq 0$; 
(\hspace{.08em}ii\hspace{.08em}) strong energy condition, $\epsilon+2\Pi\geq 0$ and $\epsilon+\Pi \geq 0$; 
(i\hspace{-.08em}i\hspace{-.08em}i) null energy condition, $\epsilon\ge 0$ and $\epsilon+\Pi \geq 0$; and 
(i\hspace{-.08em}v\hspace{-.06em}) dominant energy condition, $\epsilon\geq |\Pi|$.
For the vacuum region (i.e., $\epsilon=0$ and $\Pi=0$), 
all these energy conditions are trivially satisfied, and thus, $m'=0$ holds.
For the nonvacuum region, Conditions~(\hspace{.18em}i\hspace{.18em})--(i\hspace{-.08em}i\hspace{-.08em}i) under the inequality~\eqref{eq:maspt} provide a common 
inequality, $m'>0$. On the other hand, Condition~(i\hspace{-.08em}v\hspace{-.06em}) can be reduced to 
$m'> 0$ and $0< m\le 2r/5$.
This means that if a black hole with mass $M_0$ is centered and a matter exists in the region $r<2.5M_0$, then the violation of 
the dominant energy condition necessarily occurs there. 
Therefore, if a matter distribution surrounding the black hole satisfies 
all the standard energy conditions, a 
vacuum region must exist between the horizon and the inner boundary of the distribution. 
Note that if any one of the energy conditions is imposed together with the assumption~\eqref{eq:maspt}, the quantities $\epsilon$ and $\Pi$ will be positive.

The Einstein cluster~\cite{Einstein:1939,Geralico:2012jt} is a 
physical model compatible with the above $T^\mu{}_\nu$. This cluster is static and spherically symmetric and consists of 
an averaged distribution of collisionless particles. 
The motion of each particle in the cluster is circular geodesic motion. 
The counterrotating particles cancel out their angular momenta, so that the spherical symmetry is recovered.
Let $n(r)$ and $L_{\mathrm{p}}(r)$ be the proper number density of counterrotating 
particles with rest mass $m_{\mathrm{p}}$ and the total angular momentum of each 
of the particles moving on a circular geodesic with radius $r$, respectively.
Then, as summarized in Appendix~\ref{sec:A}, the stress-energy tensor $T^\mu{}_\nu$ is free from radial pressure and coincides with Eq.~\eqref{eq:Tmunu}, where
\begin{align}
\label{eq:ep2}
\epsilon&=m_{\mathrm{p}} n(r) \:\!\left(
1+\frac{l_{\mathrm{p}}^2}{r^2}
\right)=m_{\mathrm{p}} n(r) \:\!\frac{r-2m}{r-3m},
\\
\label{eq:PI2}
\Pi&
=m_{\mathrm{p}} n(r) \:\!
\frac{l_{\mathrm{p}}^2}{2r^2}
=\frac{1}{2} 
\frac{l_{\mathrm{p}}^2}{r^2+l_{\mathrm{p}}^2}
\epsilon(r)=m_{\mathrm{p}} n(r) \:\!\frac{m}{2(r-3m)},
\end{align}
where $l_{\mathrm{p}}=L_{\mathrm{p}}/m_{\mathrm{p}}$.
These expressions imply that $\epsilon\geq 0$ and $\Pi\geq 0$, and therefore, $m$ is further restricted as
\begin{align}
\label{eq:ECcond}
0\le m<\frac{r}{3} \quad \mathrm{and} \quad 
m'\geq0.
\end{align}
Furthermore, 
these restrictions with Eq.~\eqref{eq:malpha} lead to 
\begin{align}
\label{eq:alphaineq}
\alpha' r \leq \alpha < (1+\alpha') \frac{r}{3} 
\quad \mathrm{and} \quad 
\alpha''\le 0,
\end{align}
where we have used the relation
\begin{align}
m'=-\frac{(r-2\alpha) \alpha''}{(1-2\alpha')^2}.
\end{align}
Therefore, we see that the Einstein cluster automatically satisfies all of the above energy conditions.
To obtain an Einstein cluster, we need to give either 
$m$ and $\alpha$ so that the inequalities~\eqref{eq:ECcond} and \eqref{eq:alphaineq} hold.

\begin{figure}[t]
\centering
\includegraphics[width=7.0cm,clip]{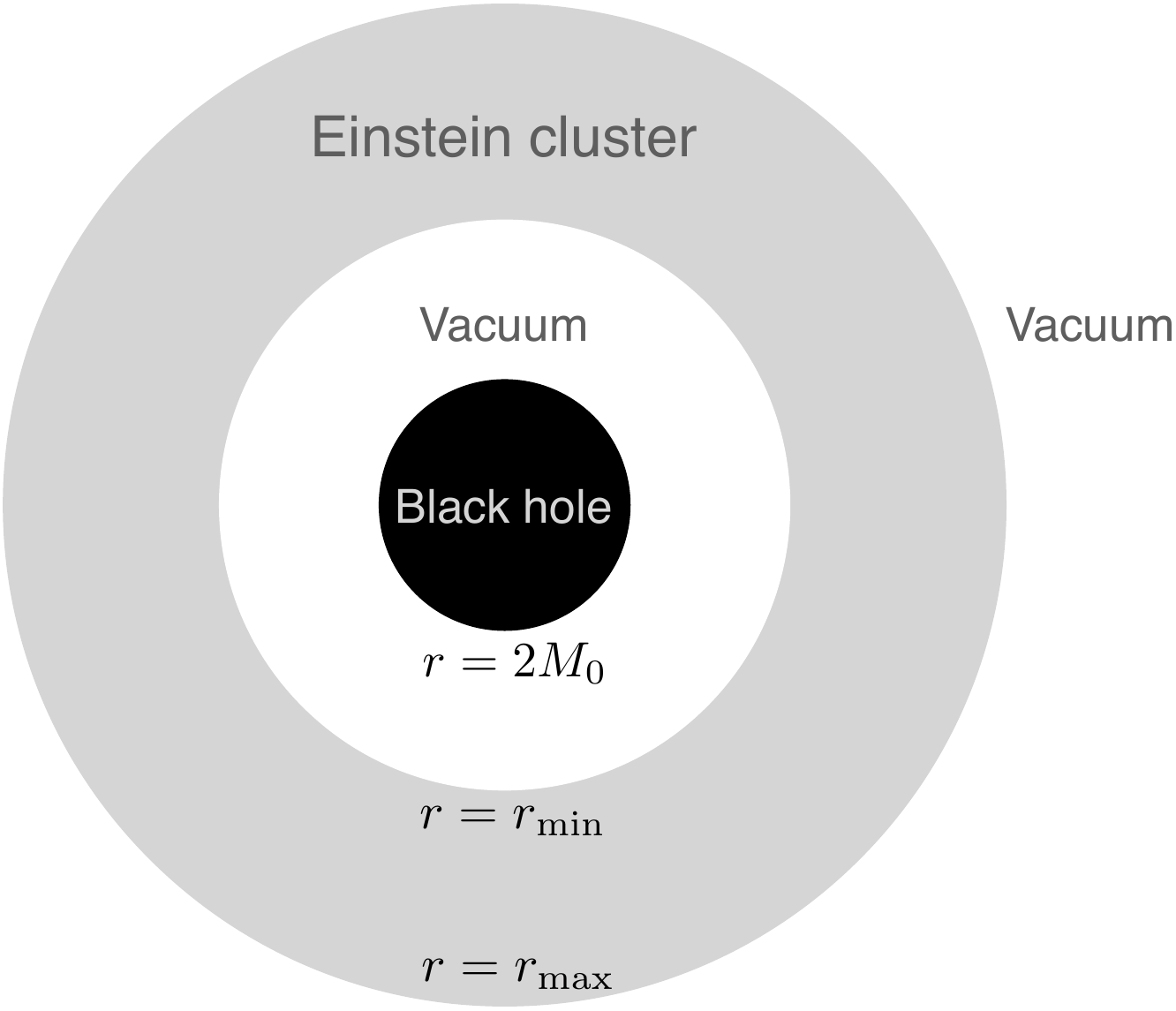}
\caption{
Schematic picture of a black hole with mass $M_0$ surrounded by an Einstein cluster distributed in the region $r_{\mathrm{min}}\le r\le r_{\mathrm{max}}$.
}
\label{fig:TS}
\end{figure}

Hereafter, we particularly focus on 
a black hole surrounded by the Einstein cluster of collisionless particles.
We assume that the mass function $m$ is of the form
\begin{empheq}[left={m=\empheqlbrace}]{alignat=4}
&
\label{eq:mnear}
M_0
\quad &\mathrm{for} \quad &2 M_0<r\le r_{\mathrm{min}},
\\
&
\label{eq:mstar}
m_*(r)
\quad &\mathrm{for} \quad &r_{\mathrm{min}}\leq r\leq r_{\mathrm{max}},
\\
&
\label{eq:mfar}
M
\quad &\mathrm{for} \quad &r\geq r_{\mathrm{max}},
\end{empheq}
where $M_0$, $M$, $r_{\mathrm{min}}$, and $r_{\mathrm{max}}$ are positive constants, 
and $m_*(r)$ is a continuous mass function of $r$ and must satisfy
\begin{align}
m_*(r_{\mathrm{min}})&=M_0,
\\
m_*(r_{\mathrm{max}})&=M. 
\end{align}
This model is known as the thick Einstein shell~\cite{Comer:1993rx}. 
Figure~\ref{fig:TS} shows a schematic picture of a black hole with mass $M_0$ surrounded by an Einstein cluster distributed in the region $r_{\mathrm{min}}\le r\le r_{\mathrm{max}}$. 
%%%%%
%%%%%
Then the corresponding $\alpha$ can be obtained by solving Eq.~\eqref{eq:alp1/2} as%
\footnote{
If we first assume the form of $m$, then 
$g_{tt}$
is restricted by the continuity of the metric at $r=r_{\mathrm{min}}$ to the following form:
\begin{align}
g_{tt}
=-\frac{r_{\mathrm{min}}-2M_0}{r}\exp \left[\:\!
\int_{r_{\mathrm{min}}}^r \frac{\mathrm{d}\tilde{r}}{\tilde{r}-2m(\tilde{r})}
\:\!\right].
\end{align}}
\begin{empheq}[left={\alpha(r)=\empheqlbrace}]{alignat=4}
&
\frac{r}{2}-\frac{C_0^2}{2}(r-2M_0)
\quad &\mathrm{for} \quad &2 M_0<r\le r_{\mathrm{min}},
\label{eq:near}
\\
&\alpha_*(r) \quad &\mathrm{for} \quad &r_{\mathrm{min}}\leq r\leq r_{\mathrm{max}},
\label{eq:matre}
\\
&
\frac{r}{2}-\frac{C^2}{2}(r-2M)
\quad &\mathrm{for} \quad &r\geq r_{\mathrm{max}},
\label{eq:far}
\end{empheq}
where $C_0$ and $C$ are integral constants, 
and $\alpha_*(r)$ is a continuous function of $r$ and must satisfy
\begin{align}
\label{eq:alpbc1}
\alpha_*(r_{\mathrm{min}})&=\frac{r_{\mathrm{min}}}{2}-\frac{C_0^2}{2}(r_{\mathrm{min}}-2M_0),
\\
\label{eq:alpbc2}
\alpha_*(r_{\mathrm{max}})&=\frac{r_{\mathrm{max}}}{2}-\frac{C^2}{2} (r_{\mathrm{max}}-2M).
\end{align}
In the region $2M_0<r\le r_{\mathrm{min}}$, the metric~\eqref{eq:met2} is reduced to the Schwarzschild spacetime with mass $M_0$, 
\begin{align}
\mathrm{d}s^2=-\left(1-2M_0/r\right)\mathrm{d}\tilde{t}^{\,2}+\left(1-2M_0/r\right)^{-1}\mathrm{d}r^2+r^2 (\mathrm{d}\theta^2+\sin^2\theta \:\!\mathrm{d}\varphi^2),
\end{align}
where $\tilde{t}=C_0 t$ is the Schwarzschild time. Thus, $M_0$ is the mass of the central black hole. In the region $r\ge r_{\mathrm{max}}$, the metric~\eqref{eq:met2} is reduced to the Schwarzschild spacetime with mass $M$, 
\begin{align}
\mathrm{d}s^2=-\left(1-2M/r\right)\mathrm{d}T^2+\left(1-2M/r\right)^{-1}\mathrm{d}r^2+r^2 (\mathrm{d}\theta^2+\sin^2\theta \:\!\mathrm{d}\varphi^2),
\end{align}
where $T=C t$ is the Schwarzschild time.
If we set $C=1$, the time $t$ is the proper time for asymptotic static observers.
Thus, $M$ is the sum of the masses of the black hole and the matter 
(i.e., the Arnowitt-Deser-Misner mass).

To obtain the distribution of an Einstein cluster, we must impose the inequalities~\eqref{eq:ECcond}, or equivalently, 
\begin{align}
0<m_*<\frac{r}{3}, \quad m'_*>0.
\end{align}
The second inequality 
$m_*<r/3$ evaluated at $r=r_{\mathrm{min}}$ and $r=r_{\mathrm{max}}$ provides lower bounds of $r_{\mathrm{min}}$ and $r_{\mathrm{max}}$, respectively,
\begin{align}
r_{\mathrm{min}}&>3M_0,
\\
r_{\mathrm{max}}&>3M.
\end{align}
Therefore, our Einstein cluster satisfying all the standard energy conditions inevitably shows a vacuum region at least in $r<3M_0$. 
Considering the stability of the Einstein clusters, we hereafter assume $r_{\mathrm{min}}>6M_0$ in almost all setups. 
We also obtain $M>M_0$ because of 
the third inequality $m'_*>0$.

%%%%%%%%%%
\section{Stellar dynamics in an Einstein cluster around a black hole}
\label{sec:3}
%%%%%%%%%%
We consider stellar
dynamics in an Einstein cluster around a black hole. 
We assume that the matter field contributes to the particle motion only through the gravitational field and that local interactions between the 
star and the 
density distribution
(e.g., pressure, friction, etc.) are negligible.
The Lagrangian of 
a massive particle
with unit mass is given by
\begin{align}
\label{eq:Lagra}
\mathscr{L}=\frac{1}{2} \left[\:\!
-\left(1-\frac{2\alpha}{r}\right) \dot{t}^2+\left(1-\frac{2m}{r}\right)^{-1} \dot{r}^2+r^2\dot{\theta}^2+r^2\sin^2\theta \:\!\dot{\varphi}^2
\:\!\right],
\end{align}
where the dot denotes differentiation with respect to proper time. 
Without loss of generality, we assume from spherical symmetry that a freely falling particle moves on the equatorial plane $\theta=\pi/2$. 
Since $t$ and $\varphi$ are cyclic variables in this mechanical system, 
the conjugate momenta 
are conserved, 
\begin{align}
\label{eq:En}
\frac{\partial \mathscr{L}}{\partial \dot{t}}&=-\left(1-\frac{2\alpha}{r}\right) \dot{t}=-E,
\\
\label{eq:L}
\frac{\partial \mathscr{L}}{\partial \dot{\varphi}}&=r^2 \dot{\varphi}=L,
\end{align}
where $E$ and $L$ are constant energy and angular momentum per unit mass of a particle, 
respectively. 
The remaining Euler-Lagrange equation for $r$ is written as
\begin{align}
\label{eq:ELeq}
&\ddot{r}+V'=0,
\\
&V(r)=\frac{1}{2}\left(1-\frac{2m}{r}\right)\left(\frac{L^2}{r^2}+1\right)-\frac{r-2m}{r-2\alpha}\frac{E^2}{2},
\end{align}
where the prime denotes differentiation with respect to $r$, and we have used Eqs.~\eqref{eq:En}, \eqref{eq:L}, and the normalization for the four-velocity, 
$\mathscr{L}=-1/2$. Integrating Eq.~\eqref{eq:ELeq}, we obtain
\begin{align}
\label{eq:constr}
\frac{1}{2}\dot{r}^2+V=0,
\end{align}
which corresponds to the normalization condition in terms of $E$ and $L$.

We focus on circular orbits, where particles must satisfy the stationary conditions
\begin{align}
\dot{r}=0 \quad \mathrm{and} \quad \ddot{r}=0. 
\end{align}
Through Eqs.~\eqref{eq:ELeq} and \eqref{eq:constr}, these conditions are rewritten as
\begin{align}
V=0 \quad \mathrm{and} \quad V'=0. 
\end{align}
Solving these algebraic
equations for $L$ and $E$, we obtain angular momentum and energy for circular orbits as functions of the orbital radius,
\begin{align}
\label{eq:Lr}
L^2(r)&=\frac{mr^2}{r-3m},
\\
\label{eq:Er}
E^2(r)&=\left(1-\frac{2\alpha}{r}\right) \frac{r-2m}{r-3m},
\end{align}
respectively. These expressions imply that the circular orbits only exist 
in the range
\begin{align}
r-3m>0.
\end{align}
The circular orbits are stable if $V''>0$, marginally (un)stable if $V''=0$, and unstable if $V''<0$, where $V''$ is the second derivative of $V$ on circular orbits given by
\begin{align}
V''
=\frac{(r-6m)m+m' r^2}{r^3(r-3m)},
\end{align}
where Eqs.~\eqref{eq:Lr} and \eqref{eq:Er} were used to eliminate $L^2$ and $E^2$, respectively.

We consider bound orbits
that are nearly circular orbits, i.e., motion of a particle which is displaced slightly from the equilibrium radius of a stable circular orbit.
For sufficiently small displacement, we can introduce two frequencies
\begin{align}
\label{eq:omegavarphi}
\omega_\varphi&= \dot{\varphi}=\sqrt{\frac{m}{r^2(r-3m)}},
\\
\label{eq:omegar}
\omega_r&=\sqrt{V''},
\end{align}
where $\omega_\varphi$ is the angular frequency, and 
$\omega_r$ is the frequency of radial harmonic oscillation.
The periapsis shift
of the nearly circular orbits 
is
measured by the precession rate 
\begin{align}
\nu&=\frac{\omega_\varphi-\omega_r}{\omega_\varphi}
\\
&=1-\sqrt{1-\frac{6m}{r}+\frac{m' r}{m}
}
\\
&=1-\sqrt{1+3\Big(\zeta-\frac{2m}{r}\Big)},
\end{align}
where $\zeta$ is the ratio of $\epsilon$ to 
the average mass density $\bar{\epsilon}(r)$ inside radius $r$,
\begin{align}
\zeta(r)&=\frac{\epsilon}{\bar{\epsilon}}=\frac{m' r}{3m}, 
\\
\bar{\epsilon}(r)&=\frac{3m}{4\pi r^3}.
\end{align}
The ratio $\zeta$ indicates the local-density effect of matters, which negatively contributes to $\nu$. 
In contrast, the ratio $2m/r$ indicates how close $r$ is to the gravitational radius $2m$ for the mass inside $r$ and can be regarded as the 
general-relativistic effect, which positively contributes to $\nu$.
We should note that $\nu$ contains $m$ and $m'$ but not $\alpha$. 
The form of $\nu$ implies that 
\begin{align}
0<\nu<1 \quad &\mathrm{for} \quad 
\zeta <\frac{2m}{r}<\zeta+\frac{1}{3},
\\
\nu=0 \quad &\mathrm{for} \quad 
\zeta=\frac{2m}{r},
\\
\nu<0 \quad &\mathrm{for} \quad 
\zeta>\frac{2m}{r}.
\end{align}
Hence, if the general-relativistic effect is larger than the local-density effect, the periapsis shift is prograde, whereas if it is smaller, the shift is retrograde.

%%%%%
\section{Periapsis shifts in specific models}
\label{sec:4}
%%%%%

We parametrize the density profile of the Einstein cluster as (see, e.g., Ref.~\cite{Hernquist:1990be})
\begin{align}
\epsilon(r)=\epsilon_* \, (r/d)^{-\gamma} \left[\:\!
1+(r/d)^\alpha
\:\!\right]^{(\gamma-\beta)/\alpha}
\Theta(r-r_{\mathrm{min}})\Theta(r_{\mathrm{max}}-r),
\end{align}
where 
$\epsilon_*$ is a constant, and 
$\Theta(\cdot)$ is the step function. 
In the following subsections, we focus on three models:  
(A) the constant density model as a simple toy model corresponding to $(\alpha, \beta, \gamma)=(\alpha, 0, 0)$, where $\alpha$ is indefinite; 
(B) the singular isothermal sphere model as the simplest dark halo model 
corresponding to 
$(\alpha, \beta, \gamma)=(\alpha, 2, 2)$, where $\alpha$ is indefinite~\cite{Binney:2008};
(C) the Navarro-Frenk-White~(NFW) profile, 
the universal profile of the dark matter distribution predicted by numerical simulations,   corresponding to 
 $(\alpha, \beta, \gamma)=(1, 3, 1)$~\cite{Navarro:1995iw}. 
The model parameters we use are summarized in Table~\ref{table:modelsummary}.

\begin{table}
\begin{tabular}{llllllllllll}
\hline
\hline
&\multicolumn{4}{c}{(A) Constant density model}&~
&\multicolumn{6}{c}{(B) Isothermal sphere model}
\\
$(\alpha, \beta, \gamma)$
&\multicolumn{4}{c}{$(\alpha, 0, 0)$}
&&\multicolumn{6}{c}{$(\alpha, 2, 2)$}
\\
\cline{2-5}
\cline{7-12}
model~~~~~~&C1~~~~~~~~&C2~~~~~~~~&C3~~~~~~~~~~&C4~~~~~~~~~~&
&I1~~~~~~~~&I2~~~~~~~~&I3~~~~~~~~&I4~~~~~~~~&I5~~~~~~~~~~&I6
\\
\cline{1-5}
\cline{7-12}
$M$
&2&1.2&2&1.2
&
&2.2&2.2&2.2&1.2
&2&2
\\
$r_{\mathrm{min}}$
&6&6&6&6
&
&5.8&6&6.2&10
&6&5
\\
$r_{\mathrm{max}}$
&$\le30$&$\le30$&30&30
&
&$\le30$&$\le30$&$\le30$&$\le30$
&10&30
\\
Figure&2(a)&2(b)&3(a)--3(c)&3(d)--3(f)
&
&4(a)&4(b)&4(c)&4(d)&5(a)--5(c)&5(d)--5(f)
\\
\hline
\hline
\end{tabular}
\\[1mm]
\begin{tabular}{lllllllll}
\hline
\hline
&\multicolumn{8}{c}{(C) NFW model}
\\
$(\alpha,\beta,\gamma)$&\multicolumn{8}{c}{$(1, 3, 1)$}
\\
\cline{2-9}
model~~~~~~~&N1~~~~~~~~&N2~~~~~~~~&N3~~~~~~~~&N4~~~~~~~~&N5~~~~~~~~&N6~~~~~~~~
&N7~~~~~~~~&N8
\\
\hline
$M$
&2&2&2&1.2&1.2&1.2
&2&2
\\
$r_{\mathrm{min}}$
&6&6&6&10&10&10
&6&6
\\
$r_{\mathrm{max}}$
&$\le30$&$\le30$&$\le30$&$\le30$&$\le30$&$\le30$
&30&30
\\
$d$
&20&9&6&100&25&5
&20&6
\\
Figure&6(a)&6(b)&6(c)&6(d)&6(e)&6(f)&7(a)--7(c)&7(d)--7(f)
\\
\hline
\hline
\end{tabular}
~~~~~~~~~~~~~~~~~~~~~~
\caption{Summary of model parameters of Einstein clusters in units where $M_0=1$.
The label C refers to the constant density model, I to the isothermal sphere model, and N to the NFW model.}
\label{table:modelsummary}
\end{table}

%%%%%
\subsection{Constant density model}
\label{sec:4A}
%%%%%
We consider the continuous mass function~\eqref{eq:mnear}--\eqref{eq:mfar} with 
\begin{align}
m_*=M_0+\frac{4\pi \epsilon_*}{3} (r^3-r_{\mathrm{min}}^3),
\end{align}
where $\epsilon_*$ is a constant given by
\begin{align}
\epsilon_*=\frac{3}{4\pi} \frac{M-M_0}{r_{\mathrm{max}}^3-r_{\mathrm{min}}^3}.
\end{align}
This mass distribution is produced by the rectangular-shaped energy density profile
\begin{align}
\epsilon=\epsilon_* \Theta(r-r_{\mathrm{min}})\Theta(r_{\mathrm{max}}-r).
\end{align}
Therefore, we call this model the constant density model.
There is no analytical expression for $\alpha$.

It is worthwhile to consider 
the innermost stable circular orbit (ISCO), which satisfies $V=0$, $V'=0$, and $V''=0$. 
Provided that the mass fraction of the cluster is sufficiently small, i.e., $\eta=(M-M_0)/M_0\ll1$,
if the ISCO appears on the matter distribution, then the radius is given by
\begin{align}
r=6M_0 \left[\:\!
1-\frac{r_{\mathrm{min}}^3+432 M_0^3}{r_{\mathrm{max}}^3-r_{\mathrm{min}}^3}
\eta
+O(\eta^2)
\:\!\right]. 
\end{align}
This means that the ISCO radius is smaller than $6M_0$ of the Schwarzschild, which is caused by the matter distribution.

\begin{figure}[t]
\centering
\includegraphics[width=12cm,clip]{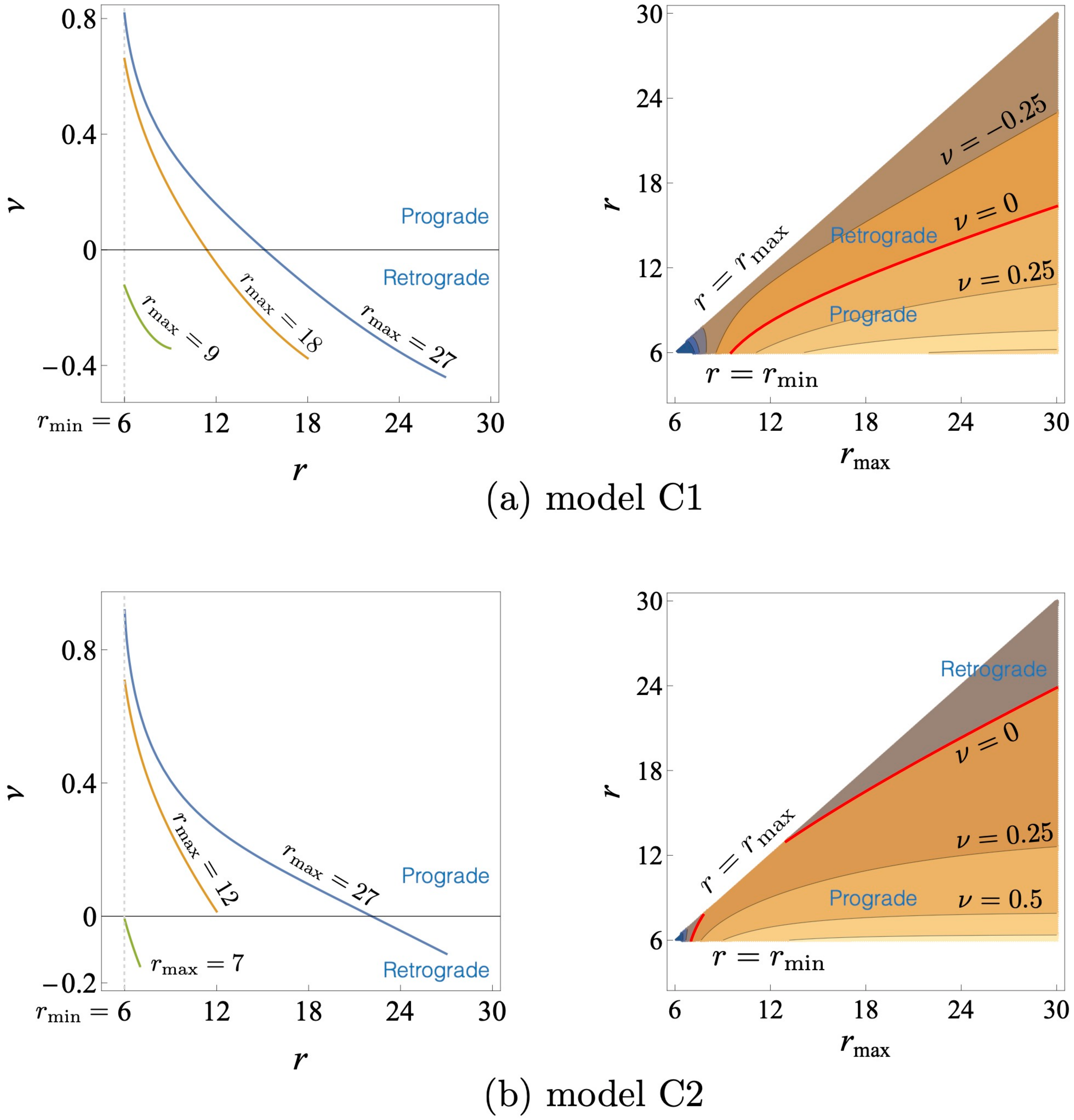}
\caption{
Precession rate $\nu$ on the matter distribution for the constant density models C1 and C2. 
See Table~\ref{table:modelsummary} for the summary of the parameter values.
The left panels show $\nu$ as a function of $r$ for several fixed values of $r_{\mathrm{max}}$.
The right panels show the contours of $\nu$ in the range $r_{\mathrm{min}}\le r\le r_{\mathrm{max}}$. 
The red solid curves denote $\nu=0$. 
}
\label{fig:cdens}
\end{figure}

We focus on the nearly circular 
bound orbits on the matter distribution. 
In this model, the two ratios $2m/r$ and $\zeta=\epsilon/\bar{\epsilon}$ are reduced to
\begin{align}
\frac{2m}{r}
=\frac{2M_0}{r}+\frac{2(M-M_0) (r^3-r_{\mathrm{min}}^3)}{r(r_{\mathrm{max}}^3-r_{\mathrm{min}}^3)},
\quad
\zeta
=\frac{(M-M_0) r^3}{(M-M_0)r^3+M_0 r_{\mathrm{max}}^3-M r_{\mathrm{min}}^3}.
\end{align}
Figure~\ref{fig:cdens}(a) shows the contour plots of $\nu$ on the matter distribution for model C1. 
If $r_{\mathrm{max}}<9.524$, then $\nu<0$. 
When $r_{\mathrm{max}}$ is relatively small, $\epsilon_*$ is relatively large. 
In this situation, the local-density effect becomes 
dominant over the 
general-relativistic effect, and as a result, retrograde shifts are more likely to occur.
On the other hand, if $r_{\mathrm{max}}> 9.524$, then 
the region of $\nu> 0$ appears near $r=r_{\mathrm{min}}$. 
When $r_{\mathrm{max}}$ is relatively large, $\epsilon_*$ is relatively small. 
Therefore, the general-relativistic effect becomes 
dominant over the local-density
effect, and as a result, prograde shifts are more likely to occur.
Figure~\ref{fig:cdens}(b) 
shows the result for model C2. 
If $r_{\mathrm{max}}<7.017$, then $\nu<0$. 
If $7.770< r_{\mathrm{max}}< 12.976$, then $\nu> 0$. 
If $7.017< r_{\mathrm{max}}<7.770$ and $r_{\mathrm{max}}> 12.976$, 
then $\nu> 0$ near $r=r_{\mathrm{min}}$ and $\nu<0$ near $r=r_{\mathrm{max}}$.
Compared to the case of Fig.~\ref{fig:cdens}(a), the region of $\nu>0$ appears from a smaller $r_{\mathrm{max}}$. 
This implies that the local-density effect is weaker,
and the general-relativistic effect tends to dominate in a wider parameter range. 
%%%%%
%%%%%

We now focus on a situation where the total mass of the matter distribution is 
much smaller than the black hole mass, $\eta=(M-M_0)/M_0\ll 1$. 
We also assume that the matter is distributed widely enough and far enough away from the black hole, 
$r_{\mathrm{max}}/M_0\gg 1$ and $(r_{\mathrm{max}}-r_{\mathrm{min}})/M_0 \gg1$. 
Then we can estimate the radius at which $\nu=0$ as 
\begin{align}
r\simeq \left(\frac{2M_0 r_{\mathrm{max}}^3}{\eta}\right)^{1/4}
\approx 480 \left(\frac{M_0}{4.0\times 10^{6} M_{\odot}}\right)^{1/4}
\left(\frac{r_{\mathrm{max}}}{1.9\times 10^3 \,\mathrm{au}}\right)^{3/4}
\left(\frac{0.01}{\eta}\right)^{1/4} \,\mathrm{au},
\end{align}
where $M_\odot$ is the solar mass, and the typical values of $M_0$ and $r_{\mathrm{max}}$ are chosen as the mass of Sgr~A$^\ast$ 
and the apoapsis distance of S2/S0-2, respectively. 
The typical value $480 \,\mathrm{au}$ 
is between the periapsis and apoapsis distances of S2/S0-2.
Thus, we see that the general-relativistic effect 
can be
canceled out by the 
local-density effect here, and the retrograde 
periapsis shift occurs outside this radius. 
It suggests that the 
local-density
effect may compensate the general-relativistic effect 
even if the matter distribution has only a mass fraction of $1\%$ relative to the black hole.

\begin{figure}[t]
\centering
\includegraphics[width=15cm,clip]{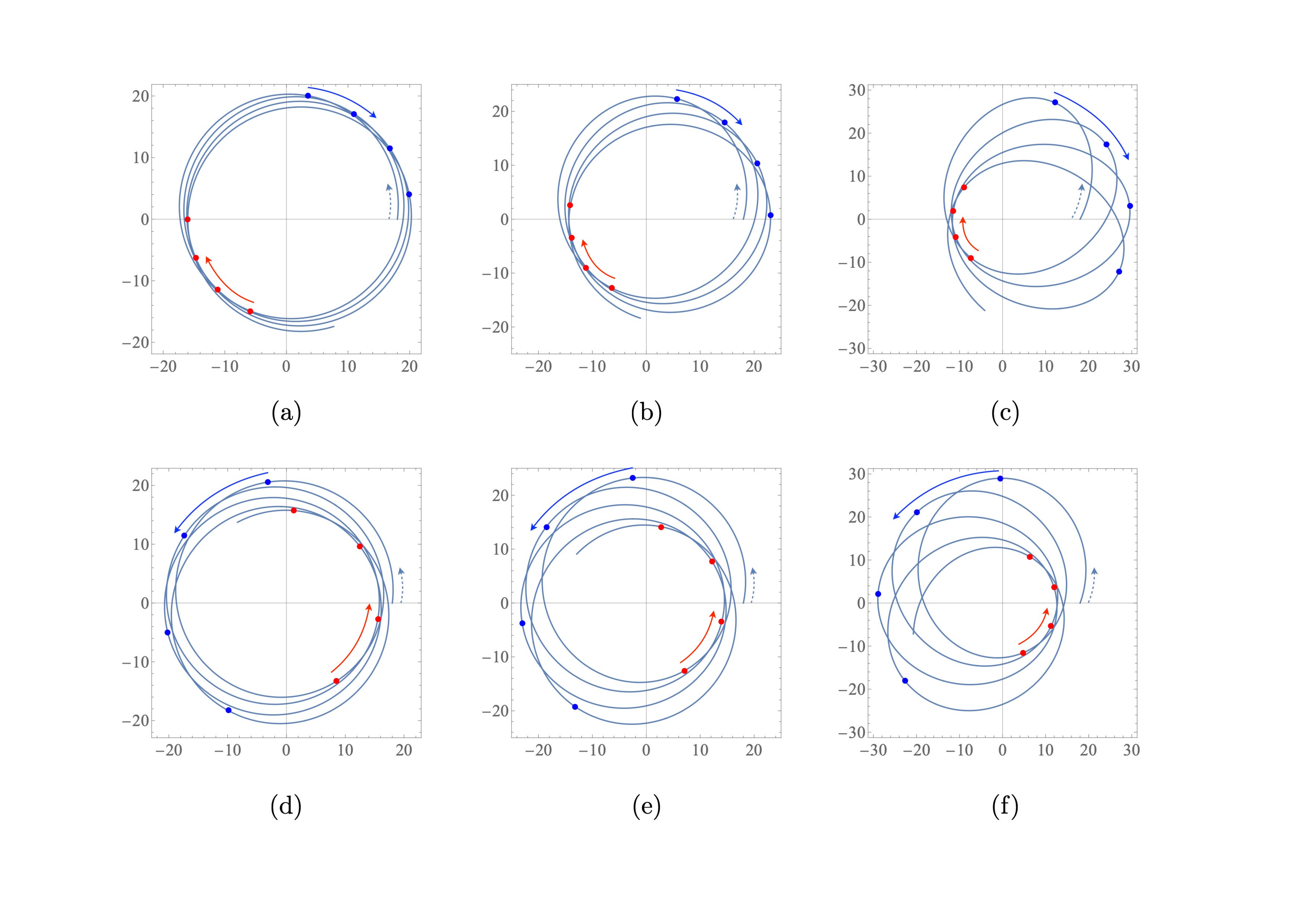}
\caption{Periapsis (red dots) and apoapsis (blue dots) shift of bound orbits (solid curves) on the x-y plane in models C3 and C4, where $(x, y)=(r \cos \varphi, r\sin \varphi)$.
See Table~\ref{table:CDMini} for initial conditions. 
}
\label{fig:cdns}
\end{figure}

Figure~\ref{fig:cdns} shows 
several bound orbits of 
stars moving in the matter distribution,
corresponding Figs.~\ref{fig:cdns}(a)--\ref{fig:cdns}(c) to model C3, and 
Figs.~\ref{fig:cdns}(d)--\ref{fig:cdns}(f) to model C4.
The solid curves denote the orbits of stars revolving counterclockwise
in the x-y plane, where $(x,y)=(r\cos \varphi, r \sin \varphi)$. 
When a small amount of energy $\Delta E$ is injected to a star on a circular orbit, 
the orbit transits to a 
bound orbit
of nearly circular shape [see Figs.~\ref{fig:cdns}(a) and \ref{fig:cdns}(d)]. 
The amplitude of radial oscillation increases with $\Delta E$ [see Figs.~\ref{fig:cdns}(b) and \ref{fig:cdns}(e)]. In Figs~\ref{fig:cdns}(c) and \ref{fig:cdns}(f), each orbit extends across the entire matter distribution. 
The red and blue dots indicate the periapsises and apoapsises of the bound orbits, respectively.
Figures~\ref{fig:cdns}(a)--\ref{fig:cdns}(c) show their retrograde shifts, 
where the blue and red dots revolve clockwise. 
On the other hand, 
Figs~\ref{fig:cdns}(d)--\ref{fig:cdns}(f) show their prograde shifts, 
where the blue and red dots revolve counterclockwise.
We can see that even if the amplitude of the radial oscillation becomes large by injecting energy, the shift direction is the same as 
in the case of nearly circular 
bound orbits.

%%%%%
\begin{table}
\begin{tabular}{lllllll}
\hline
\hline
Figure~~~~~~~~~
&3(a)~~~~~~~~~~~
&3(b)~~~~~~~~~~~
&3(c)~~~~~~~~~~~
&3(d)~~~~~~~~~~~
&3(e)~~~~~~~~~~~
&3(f)
\\
\hline
$r(0) (=r_{\mathrm{c}})$
&18&18&18&18&18&18
\\
$\varphi(0)$
&$0$
&$0$
&$0$
&$0$
&$0$
&$0$
\\
$E$
&9.934223e-1
&9.3627e-1
&9.445e-1
&9.27945e-1
&9.2896e-1
&9.3199e-1
\\
$E-E(r_{\mathrm{c}})$
&$6.82\textrm{e-}4$
&$2.73\textrm{e-}3$
&$1.09\textrm{e-}2$
&$5.05\textrm{e-}4$
&$1.52\textrm{e-}3$
&$4.55\textrm{e-}3$
\\
$L\:\! (=L(r_{\mathrm{c}}))$
&5.222
&5.222
&5.222
&5.222
&5.222
&5.222
\\
halo model &C3&C3&C3&C4&C4&C4
\\
\hline
\hline
\end{tabular}
\caption{Initial conditions of particle orbits in constant density models in units and the gauge where $M_0=1$ and $C=1$. The values of $\dot{r}(0)$ are determined from Eq.~\eqref{eq:constr}.}
\label{table:CDMini}
\end{table}
%%%%%

%%%%%
\subsection{Isothermal sphere model}
%%%%%
We consider the continuous mass function~\eqref{eq:mnear}--\eqref{eq:mfar} with
\begin{align}
m_*
&=\sigma r+ \delta,
\end{align}
where 
\begin{align}
\sigma&=\frac{M-M_0}{r_{\mathrm{max}}-r_{\mathrm{min}}},\\
\delta&=\frac{M_0 r_{\mathrm{max}}-M r_{\mathrm{min}}}{r_{\mathrm{max}}-r_{\mathrm{min}}}.
\end{align}
The mass distribution $m_*$ is produced by the density profile of the truncated singular isothermal sphere, 
\begin{align}
\epsilon=\frac{\sigma}{4\pi r^2}\Theta(r-r_{\mathrm{min}})\Theta(r_{\mathrm{max}}-r).
\end{align}
Therefore, we call this model the isothermal sphere model.
If $\sigma\neq 1/2$, the corresponding $\alpha$
is given by Eqs.~\eqref{eq:near}--\eqref{eq:far} with 
\begin{align}
\alpha_*&=\frac{r}{2}-\frac{C_*^2}{2}(r-2m_*)^{1/(1-2\sigma)},
\end{align}
where $C_*$ is an integral constant giving the shift degree of freedom of the time coordinate.
The boundary conditions~\eqref{eq:alpbc1} and \eqref{eq:alpbc2} give the relations, 
\begin{align}
C_0&=C_* (r_{\mathrm{min}}-2M_0)^{\sigma/(1-2\sigma)},
\\
C&=C_*(r_{\mathrm{max}}-2M)^{\sigma/(1-2\sigma)}.
\end{align}
If $\sigma=1/2$, the corresponding $\alpha_*$ takes the form
\begin{align}
\alpha_*=\frac{r}{2}-\frac{C_*^2}{2} e^{r/(r_{\mathrm{min}}-2M_0)},
\end{align}
where $C_*$ is an integration constant giving the shift degree of freedom of the time coordinate. 
The boundary conditions~\eqref{eq:alpbc1} and \eqref{eq:alpbc2} lead to 
\begin{align}
C_0&=C_*\:\! \frac{e^{r_{\mathrm{min}}/2(r_{\mathrm{min}}-2M_0)}}{\sqrt{r_{\mathrm{min}}-2M_0}},
\\
C
&=C_*\:\!\frac{e^{r_{\mathrm{max}}/2(r_{\mathrm{max}}-2M)}}{\sqrt{r_{\mathrm{max}}-2M}}. 
\end{align}

For 
$\eta=(M-M_0)/M_0\ll1$, 
if the ISCO appears on the matter distribution, its radius is given by
\begin{align}
r=6 M_0 \left[\:\!
1 -\frac{r_{\mathrm{min}}}{r_{\mathrm{max}}-r_{\mathrm{min}}}
\eta
+O\left(\eta^2\right)
 \:\!\right].
\end{align}
This means that the ISCO is smaller than $6M_0$ of the Schwarzschild, which is caused by the matter distribution.

We focus on the nearly circular 
bound orbits on the matter distribution.
The two ratios $2m/r$ and $\zeta$ of this model are reduced to
\begin{align}
\frac{2m}{r}&=2\:\!\sigma+\frac{2\delta}{r},
\quad
\zeta=\frac{\sigma r}{3(\sigma r+\delta)}.
\end{align}
%%%%%
\begin{figure}[t]
\centering
\includegraphics[width=11.5cm,clip]{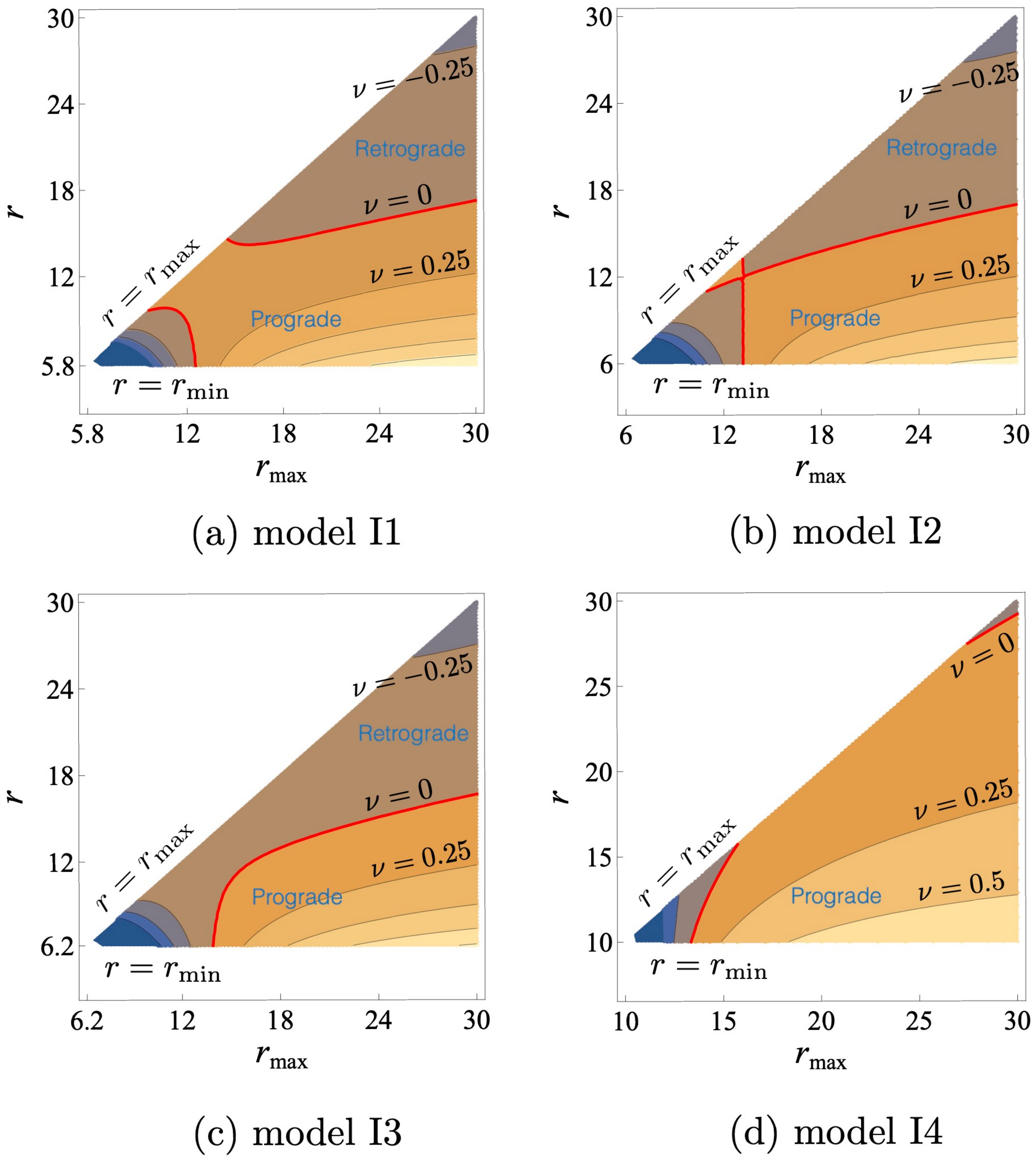}
\caption{
Contours of $\nu$ (0.25 intervals) in models I1--I4. 
See Table~\ref{table:modelsummary} for the summary of the parameter values. 
The red solid curves denote $\nu=0$. 
}
\label{fig:iso}
\end{figure}
%%%%%
Figure~\ref{fig:iso}(a) 
shows the contour plots of $\nu$ on the matter distribution in model I1. 
If $r_{\mathrm{max}}<9.640$, then $\nu<0$;
if $12.528%(exact value)
<r_{\mathrm{max}}<14.559$, 
then $\nu>0$; 
if $r_{\mathrm{max}}>14.559$, then $\nu>0$ near $r=r_{\mathrm{min}}$ and $\nu<0$ near $r=r_{\mathrm{max}}$. 
These behaviors can be interpreted in the same way as the constant density model. 
However, there is a novel 
situation, where
$\nu<0$ near $r=r_{\mathrm{min}}$ and $\nu>0$ near $r=r_{\mathrm{max}}$ in $9.640<r_{\mathrm{max}}<12.528$, 
which is not found in the constant density model (see Fig.~\ref{fig:cdens}).
Since the local density is proportional to $r^{-2}$, we can see that the 
local-density effect in this range decreases to such an extent that the 
general-relativistic effect dominates as $r$ increases. 
In Fig.~\ref{fig:iso}(b), 
the contour of $\nu=0$ makes a vertical segment at $r_{\mathrm{max}}=13.2$, 
where 
$\nu=0$ appears in the whole range of $r$. 
This is a special case unique to the isothermal sphere model.
In Fig.~\ref{fig:iso}(c), 
the plot shows qualitatively the same behavior as the case in Fig.~\ref{fig:cdens}(a). 
Figures~\ref{fig:iso}(a)--\ref{fig:iso}(c) 
show the change of the contours 
as the value of $r_{\mathrm{min}}$ gradually increases. 
Figure~\ref{fig:iso}(d) is the case of model I4 and shows
qualitatively the same behavior as the case in Fig.~\ref{fig:cdens}(b).

We now focus on a situation where $\eta\ll 1$. 
We also assume that $r_{\mathrm{max}}/M_0\gg 1$ and $(r_{\mathrm{max}}-r_{\mathrm{min}})/M_0 \gg1$. 
Then we can estimate the radius at which $\nu=0$ as 
\begin{align}
r\simeq 
\left(\frac{6 r_{\mathrm{max}} M_0}{\eta}\right)^{1/2}
\approx 210 
\left(\frac{r_{\mathrm{max}}}{1.9\times 10^3 \,\mathrm{au}}\right)^{1/2} 
\left(\frac{M_0}{4.0\times 10^{6} M_{\odot}}\right)^{1/2}\left(\frac{0.01}{\eta}\right)^{1/2}
\,\mathrm{au},
\end{align}
where the typical values are chosen as the same as in the previous subsection. 
The typical value $210 \,\mathrm{au}$ is comparable to the periapsis distance $120 \,\mathrm{au}$ of S2/S0-2.
As in the previous model, it is suggested that the local-density effect cancels out the general-relativistic effect even if the matter distribution has only a $1\%$ mass fraction relative to the black hole.

\begin{figure}[t]
\centering
\includegraphics[width=15cm,clip]{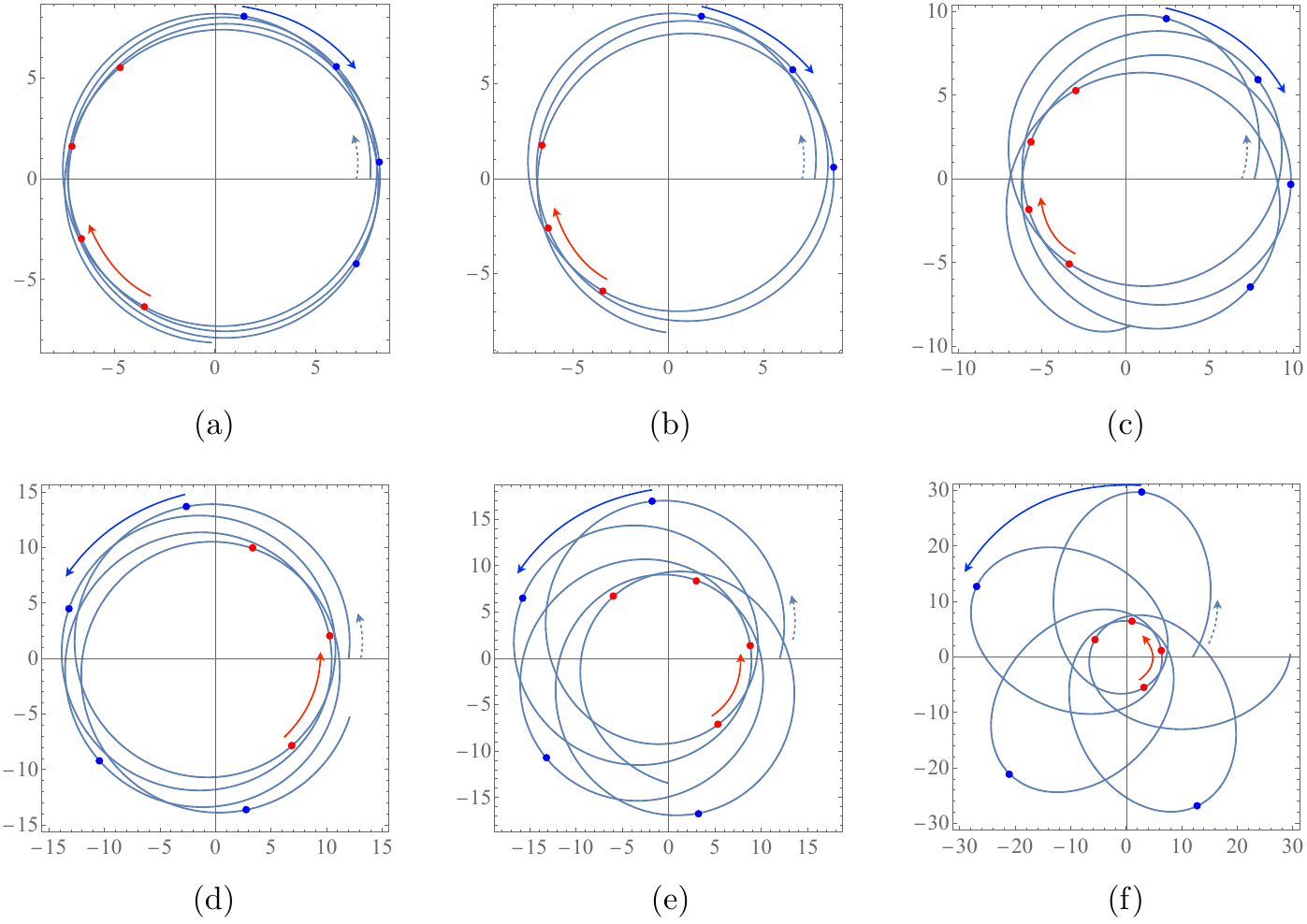}
\caption{
Periapsis (red dots) and apoapsis (blue dots) shift of bound orbits (solid curves) on the x-y plane in models I5 and I6. See Table~\ref{table:ISOini} for initial conditions.
}
\label{fig:nonliniso}
\end{figure}

Figures~\ref{fig:nonliniso}(a)--\ref{fig:nonliniso}(c) show particle orbits in model I5. 
We see that the retrograde 
periapsis shift 
occurs.
%%%%%
\begin{table}
\begin{tabular}{lllllll}
\hline
\hline
Figure~~~~~~~~~
&5(a)~~~~~~~~~~~
&5(b)~~~~~~~~~~~
&5(c)~~~~~~~~~~~
&5(d)~~~~~~~~~~~
&5(e)~~~~~~~~~~~
&5(f)
\\
\hline
$r(0)$&7.7&7.7&7.7&12&12&12
\\
$\varphi(0)$&0&0&0&0&0&0
\\
$E$&8.50007e-1&8.5271e-1&8.635e-1&9.1452e-1&9.1921e-1&9.427e-1
\\
$E-E(r_{\mathrm{c}})$&9.02e-4&3.61e-3&1.44e-2&1.17e-3&5.87e-3&2.93e-2
\\
$L\:\!(=L(r_{\mathrm{c}}))$&4.967&4.967&4.967&4.753&4.753&4.753
\\
halo model&I5&I5&I5&I6&I6&I6
\\
\hline\hline
\end{tabular}
\caption{Initial conditions of particle orbits in isothermal sphere models in units and the gauge where $M_0=1$ and $C=1$. The values of $\dot{r}(0)$ are determined from Eq.~\eqref{eq:constr}.}
\label{table:ISOini}
\end{table}
%%%%%
Figures~\ref{fig:nonliniso}(d)--\ref{fig:nonliniso}(f) show the case of model I6. 
When the energy slightly increases
from that of the circular orbit, 
the orbital shape becomes 
quasi-circular
[see Fig.~\ref{fig:nonliniso}(d)]. 
As the energy increases, 
the radial amplitude increases [see Figs.~\ref{fig:nonliniso}(e) and \ref{fig:nonliniso}(f)].
These cases show the prograde 
periapsis shifts. 
We can see that even if the amplitude of the radial oscillation becomes large by injecting energy, the shift direction is the same as 
in the 
quasi-circular
case.

%%%%%%%%%%
\subsection{NFW model}
%%%%%%%%%%
We consider the continuous mass function~\eqref{eq:mnear}--\eqref{eq:mfar} with
\begin{align}
m_*=M_0+4\pi \epsilon_{\mathrm{s}} d^3 \left[\:\!
\ln \left(
\frac{1+r/d}{1+r_{\mathrm{min}}/d}
\right)
+\frac{1}{1+r/d}-\frac{1}{1+r_{\mathrm{min}}/d}\:\!\right],
\end{align}
where $d$ is a constant called the scale radius, 
and $\epsilon_*$ is a constant given by
\begin{align}
\epsilon_* =\frac{M-M_0}{4\pi d^3}\left[\:\!
\ln \left(
\frac{1+r_{\mathrm{max}}/d}{1+r_{\mathrm{min}}/d}
\right)
+\frac{1}{1+r_{\mathrm{max}}/d}
-\frac{1}{1+r_{\mathrm{min}}/d}
\:\!\right]^{-1}.
\end{align}
The mass distribution $m_*$ is produced by the 
Navarro-Frenk-White~(NFW) profile~\cite{Navarro:1995iw}
\begin{align}
\epsilon =\frac{\epsilon_*}{(r/d)(1+r/d)^2}
\Theta(r-r_{\mathrm{min}})\Theta(r_{\mathrm{max}}-r).
\end{align}
Therefore, we call this model the NFW model.

For $\eta\ll 1$, if the ISCO appears on the matter distribution, its radius is given by
\begin{align}
r&=6M_0 \left[\:\!
1-\frac{1-\ln (6M_0/r_{\mathrm{min}})}{\ln (r_{\mathrm{max}}/r_{\mathrm{min}})} 
\eta
+O(\eta^2)
\:\!\right]
\end{align}
in the limit $d\to 0$, and 
\begin{align}
r&=6M_0 \left[\:\!
1-\frac{r_{\mathrm{min}}^2+36 M_0^2}{r_{\mathrm{max}}^2-r_{\mathrm{min}}^2}
\eta+O(\eta^2)
\:\!\right]
\end{align}
in the limit $d/M_0\to \infty$.
These values are smaller than $6M_0$ of the Schwarzschild, which is caused by the matter distribution.

\begin{figure}[t]
\centering
\includegraphics[width=15cm,clip]{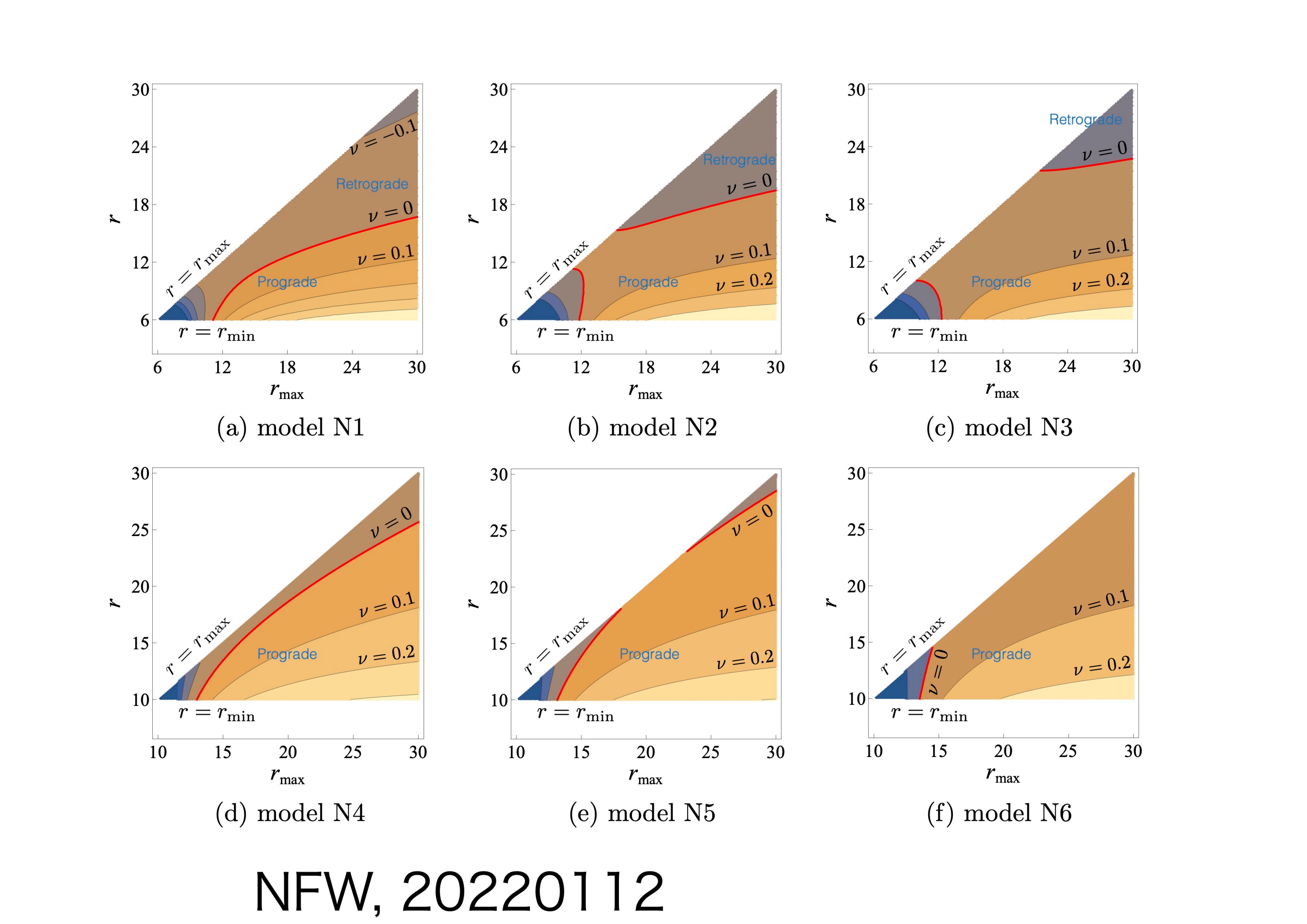}
\caption{Contour plots of 
the precession rate $\nu$ for models N1--N6. 
See Table~\ref{table:modelsummary} for the summary of the parameter values.
The red curves denote $\nu=0$. The contour interval is $0.1$.
}
\label{fig:NFWcontour}
\end{figure}

Figures~\ref{fig:NFWcontour}(a)--\ref{fig:NFWcontour}(f) show the contour plots of $\nu$ on the matter distribution in models N1--N6, respectively.  
Figures~\ref{fig:NFWcontour}(a) and \ref{fig:NFWcontour}(d) show qualitatively the same behaviors as in Figs.~\ref{fig:cdens}(a) and \ref{fig:iso}(c). 
Figure~\ref{fig:NFWcontour}(c) shows qualitatively the same behavior as in Fig.~\ref{fig:iso}(a).
Figures~\ref{fig:NFWcontour}(e) and \ref{fig:NFWcontour}(f) show qualitatively the same behaviors as in Figs.~\ref{fig:cdens}(b) and \ref{fig:iso}(d). 
Consider Fig.~\ref{fig:NFWcontour}(b), which shows a novel $\nu=0$ behavior that has not appeared in the previous cases. 
If $r_{\mathrm{max}}<11.290$, then $\nu<0$; 
if $12.217<\nu<15.324$, then $\nu>0$; 
if $\nu>15.324$, then $\nu<0$ near $r=r_{\mathrm{max}}$ and $\nu>0$ near $r=r_{\mathrm{min}}$. 
These behaviors can be interpreted 
in the same way as the constant density model and the isothermal sphere model.
If $11.290<r_{\mathrm{max}}<11.814$, then $\nu<0$ near $r=r_{\mathrm{min}}$ and $\nu>0$ near $r=r_{\mathrm{max}}$. 
The same behavior is seen in the case of Fig.~\ref{fig:iso}(a) of 
the isothermal sphere model.
If $11.814<r_{\mathrm{max}}<12.217$, 
then $\nu>0$ near $r=r_{\mathrm{min}}$ and $r=r_{\mathrm{max}}$, 
whereas $\nu<0$ in the intermediate region, where $d$ is comparable to the distribution scale. 
This behavior is not seen in the previous two models.

Assume that $M_0=4.0\times 10^{6} M_{\odot}$, 
$\eta=0.01$,
$r_{\mathrm{max}}=1.9\times 10^{3} \,\mathrm{au}$,
and 
$r_{\mathrm{min}}=1.2\times 10^2 \,\mathrm{au}$. 
Then we can estimate the radius at which $\nu=0$ as 
$r\approx 130 \,\mathrm{au}$ for $d\approx 0$, 
$r\approx 200 \,\mathrm{au}$ for $d=100 \,\mathrm{au}$, 
and
$r\approx 420 \,\mathrm{au}$ for $d=1\times 10^4 \,\mathrm{au}$. 
As in the previous two cases, the 
local-density
effect can compensate for the 
general-relativistic effect in a realistic observational range, even if the mass fraction is $1\%$.

%%%%%
%%%%%

\begin{figure}[t]
\centering
\includegraphics[width=15cm,clip]{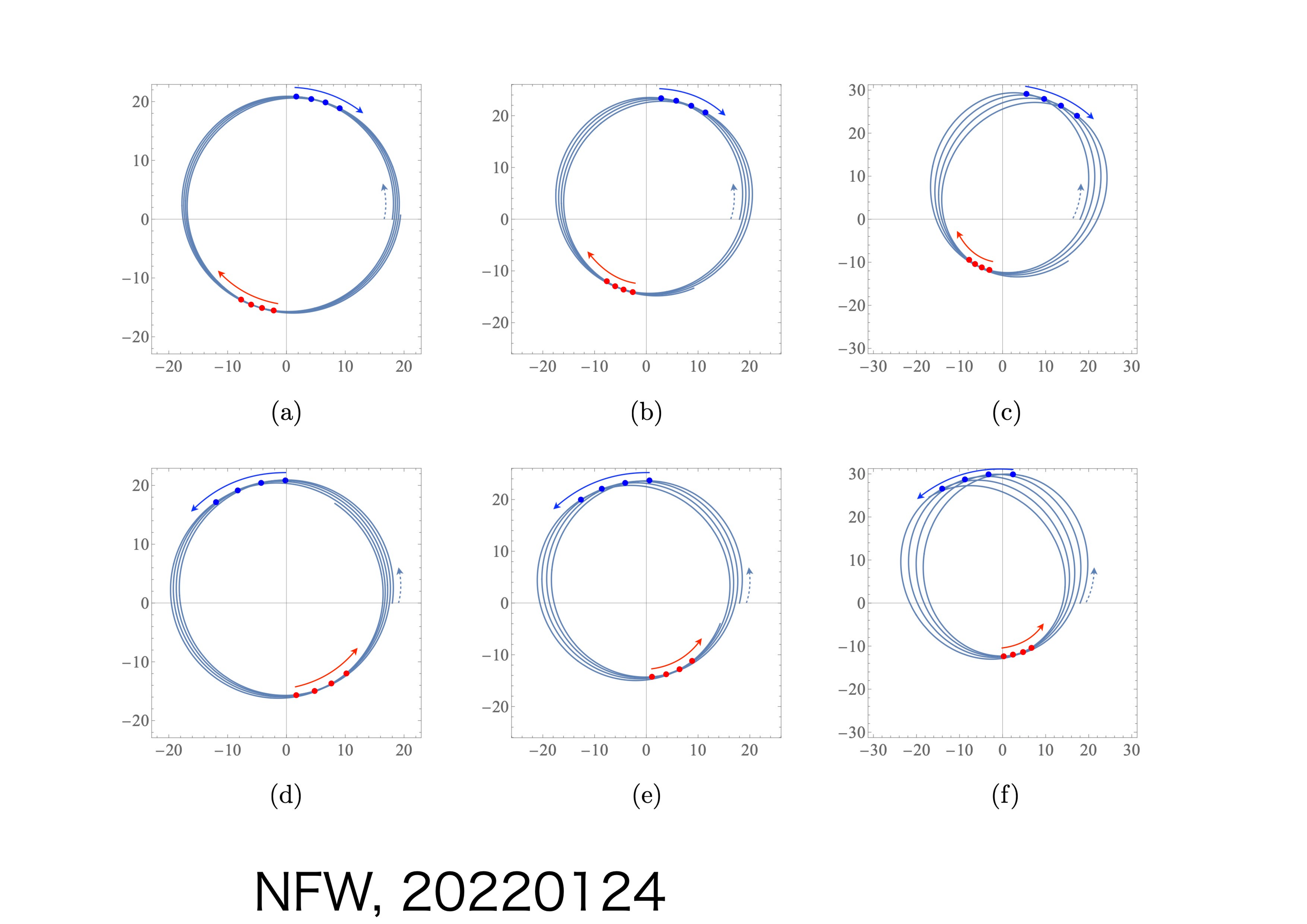}
\caption{
Bound orbits and their periapsis and apoapsis shifts in models N7 and N8.
The curves and dots are defined 
in the same way
as in Fig.~\ref{fig:cdns}.
See Table~\ref{table:NFWini} for initial conditions. 
}
\label{fig:NFWorbit}
\end{figure}

Figure~\ref{fig:NFWorbit} shows 
several bound
orbits on the matter distribution in models N7 and N8. 
As $E-E(r_{\mathrm{c}})$ increases from (a) to (c) or from (d) to (f), 
the amplitude of the radial oscillation increases.
The first line shows retrograde 
periapsis shifts, while 
the second line shows prograde 
periapsis shifts.

%%%%%
\begin{table}
\begin{tabular}{lllllll}
\hline\hline
Figure~~~~~~~~~
&7(a)~~~~~~~~~
&7(b)~~~~~~~~~
&7(c)~~~~~~~~~
&7(d)~~~~~~~~~
&7(e)~~~~~~~~~
&7(f)
\\
\hline
$r(0) \:\!(=r_{\mathrm{c}})$&18&18&18&18&18&18
\\
$\varphi(0)$&0&0&0&0&0&0
\\
$E$&9.3913e-1&9.4149e-1&9.486e-1&9.4332e-1&9.4569e-1&9.528e-1
\\
$E-E(r_{\mathrm{c}})$&1.18e-3&3.55e-3&1.06e-2&1.18e-3&3.55e-3&1.06e-2
\\
$L \:\!(=L(r_{\mathrm{c}}))$&5.946&5.946&5.946&5.946&5.946&5.946
\\
halo model&N7&N7&N7&N8&N8&N8
\\
\hline\hline
\end{tabular}
\caption{Initial conditions of particle orbits in NFW models in units and the gauge where $M_0=1$ and $C=1$. 
The values of $\dot{r}(0)$ are determined from Eq.~\eqref{eq:constr}.}
\label{table:NFWini}
\end{table}
%%%%%

%%%%%
\section{Model functions and observables}
\label{sec:5}
%%%%%
We consider whether the model functions $m(r)$, $\alpha(r)$, and $\epsilon(r)$ are determined by observations of photons coming from orbiting stars. Assume
that a distant observer observes a nearly circular 
bound
orbit from its orbital axis (i.e., face on to the orbital plane). 
Then we focus on the following four observables: 
The orbital period $T_\varphi$, 
the periapsis
shift angle $\Delta \varphi_{\mathrm{p}}$ of 
the nearly circular 
bound orbit,
the redshift factor $z$ of photons coming from the 
star (the so-called spectroscopic observable), 
and 
the angular radius $\beta$ of the orbital radius of the 
star
on the celestial sphere. 
The two observables $T_\varphi$ and $\Delta \varphi_{\mathrm{p}}$ can be measured by continuous observation at least for one orbital period. 
In the present spacetime, 
using Eqs.~\eqref{eq:En}, \eqref{eq:Er}, \eqref{eq:omegavarphi}, and \eqref{eq:omegar},
these
observables are represented as
\begin{align}
T_\varphi
&=\frac{2\pi}{\mathrm{d}\varphi/\mathrm{d}T}
=2\pi 
\sqrt{\frac{r^3}{m}\frac{r-2m}{r-2\alpha}}\bigg|_{\mathrm{s}},
\\
\Delta \varphi_{\mathrm{p}}
&=\frac{2\pi}{\omega_r}(\omega_\varphi-\omega_r)
=2\pi \frac{\nu}{1-\nu}=2\pi \left[\:\
\left(1+\frac{4\pi r^3\epsilon}{m}-\frac{6m}{r}\right)^{-1/2}-1
\:\!\right]\!\!\Bigg|_{\mathrm{s}},
\end{align}
where 
$\mathrm{d}\varphi/\mathrm{d}T=(\mathrm{d}\varphi/\mathrm{d}t)(\mathrm{d} t/\mathrm{d}T)=C^{-1} \dot{\varphi}/\dot{t}$, 
we have chosen $C=1$, 
and the symbol ``$\mathrm{s}$" means 
the quantities for the source.
The others, $z$ and $\beta$, are given by the momentum $k_\mu$ of photons coming from the star,
\begin{align}
1+z&=\frac{(k_\mu u_{\mathrm{s}}^\mu)\big|_{\mathrm{s}}}{(k_\mu u_{\mathrm{o}}^\mu)\big|_{\mathrm{o}}}=\dot{t}-b \:\!\dot{\varphi},
\\
\sin \beta&=
\sqrt{(k^{(2)}/k^{(0)})^2+(k^{(3)}/k^{(0)})^2}\Big|_{\mathrm{o}}
=\frac{q}{r_{\mathrm{o}}} \sqrt{1-\frac{2\alpha(r_{\mathrm{o}})}{r_{\mathrm{o}}}},
\end{align}
where the symbol ``$\mathrm{o}$" means 
the quantities for the observer,
and
$b$ and $q$ are constants of photon motion 
known as the impact parameters, 
\begin{align}
\label{eq:bqdef1}
b&=\frac{k_\varphi}{(-k_t)}, 
\\
\label{eq:bqdef2}
q&=\frac{1}{(-k_t)}\sqrt{k_\theta^2+\frac{k_\varphi^2}{\sin^2\theta}},
\end{align}
and
$u^{\mu}_{\mathrm{s}}=(\dot{t},0,0,\dot{\varphi})$ and $u_{\mathrm{o}}^\mu=(1,0,0,0)$ are the velocity of the 
star and that of the distant observer in the coordinates $(t,r,\theta,\varphi)$, respectively, and $k^{(\mu)}$ is the tetrad components of $k_\mu$ (see Appendix~\ref{sec:B} for details).
Using the assumption of the observer being face-on, we can set $b=0$ by the coordinate transformation, and then 
\begin{align}
\label{eq:zobs}
1+z
&
=\sqrt{\frac{r}{r-3m} \frac{r-2m}{r-2\alpha}}\bigg|_{\mathrm{s}},
\end{align}
where we have used Eqs.~\eqref{eq:En} and \eqref{eq:Er}.
Furthermore, in the face-on case, because only the photons with zero radial velocity at the emission point will reach the distant observer on the axis, the parameter $q$ takes the value 
\begin{align}
\label{eq:qsour}
q^2
=\frac{r^3}{r-2\alpha}\bigg|_{\mathrm{s}},
\end{align}
where we have used the radial equation for photon motion, 
\begin{align}
\left(\frac{\mathrm{d}r}{\mathrm{d}\lambda}\right)^2+\left(1-\frac{2m}{r}\right)\left(\frac{q^2}{r^2}-\frac{r}{r-2\alpha}\right)=0,
\end{align}
where $\lambda$ is an affine parameter. Using $r_{\mathrm{o}}\gg M$ and $\beta\ll 1$, 
we obtain 
\begin{align}
\beta&= 
\frac{r_{\mathrm{s}}}{r_{\mathrm{o}}\sqrt{
1-2\alpha(r_{\mathrm{s}})/r_{\mathrm{s}}}}. 
\end{align}
If we know the value of $r_{\mathrm{o}}$, by measuring the four observables $\Delta \varphi_{\mathrm{p}}$, $T_\varphi$, $z$, and $\beta$ for a nearly circular bound orbit in the face-on case, we obtain the values of $m(r_{\mathrm{s}})$, $\alpha(r_{\mathrm{s}})$, $\epsilon(r_{\mathrm{s}})$, and the equilibrium radius $r_{\mathrm{s}}$.
Therefore, by using observational data for several stars orbiting at different radii, 
we can 
obtain
the functions of our model 
by appropriate interpolation.
In contrast, if we do not know the value of $r_{\mathrm{o}}$, we obtain a relationship between two sets of dimensionless quantities
($\Delta \varphi_{\mathrm{p}}$, $T_{\varphi}/r_{\mathrm{s}}$ $z$, $\beta$) and ($m(r_{\mathrm{s}})/r_{\mathrm{s}}$, $\alpha(r_{\mathrm{s}})/r_{\mathrm{s}}$, $\epsilon(r_{\mathrm{s}}) r_{\mathrm{s}}^2$, and $r_{\mathrm{o}}/r_{\mathrm{s}}$).
Note that, however, 
for the observer not in the face-on to the nearly circular orbital plane of the 
star,
the situation becomes much more convoluted
due to the uncertainties in measuring the values of the orbital shift angle $\Delta\varphi_{\mathrm{p}}$ and the 
the position on the celestial sphere
$(X,Y)$ (see Appendix~\ref{sec:B}).%
\footnote{These uncertainties arise from the gravitational lens effect on the orbit of photons coming from the source to the observer. 
For $\Delta \varphi_{\mathrm{p}}$, the gravitational lens effect deforms the visible shape of 
the star orbit,
and the exact value of $\Delta \varphi_{\mathrm{p}}$ cannot be measured from the shape without prior knowledge of the metric functions $m(r)$ and $\alpha(r)$. For $(X,Y)$, although they are related to $(q, b/\sin\theta_{\mathrm{o}})$ through Eqs.~\eqref{eq:Xapp} and \eqref{eq:Yapp}, neither the 
relation
$b=0$ nor Eq.~\eqref{eq:qsour} is valid in general because of the lens effect. 
However, if the deviation from the face-on case is small, or equivalently, if the inclination angle is small 
(i.e., $i=\theta_0\ll 1$), 
the discussion for the face-on case holds approximately. 
Then the correction is typically given by the projection of the line-of-sight direction to the face-on direction, whose order is expected to be $O(1-\cos i)=O(i^2)$.}

To make the discussion clear, we focus for a while on a quasi-Newtonian nearly circular 
quasi-elliptical
orbit with an equilibrium radius $r_{\mathrm{s}}$, where $r_{\mathrm{s}}\gg m(r_{\mathrm{s}})$, $r_{\mathrm{s}}\gg \alpha(r_{\mathrm{s}})$, and $|\nu|\ll 1$. 
In this case, the periapsis shift $\Delta \varphi_{\mathrm{p}}$ is given by the following simple formula:
\begin{align}
\label{eq:shiftNe}
\Delta \varphi_{\mathrm{p}}=3\pi \left(
\frac{2m}{r}-\frac{\epsilon}{\bar{\epsilon}}
\right)\!\bigg|_{\mathrm{s}}.
\end{align}
Then we derive the concrete expressions for $m(r_{\mathrm{s}})$, $\alpha(r_{\mathrm{s}})$, and $\epsilon(r_{\mathrm{s}})$ and therefore the deviation from the Schwarzschild solution. 
For such an orbit, we can determine 
the mass $m_{\mathrm{s}}$,
the radius $r_{\mathrm{s}}$, and the inclination angle $i=\theta_{\mathrm{o}}$ using the maximum and minimum separations from the central object, 
$r_{\mathrm{s}}=r_{\mathrm{o}} \max(\beta)$ and 
$r_{\mathrm{s}} \cos i=r_{\mathrm{o}} \min(\beta)$,
respectively; 
the orbital period, $T_\varphi=2\pi \sqrt{r^3/m}|_{\mathrm{s}}$; and 
the difference between the maximum and minimum redshifts, 
\begin{align}
\max(z)-\min(z)=2 \sqrt{m/r}|_{\mathrm{s}}\sin i.
\end{align}
Then, 
if the periapsis shift angle $\Delta \varphi_{\mathrm{p}}$ is observed, 
the values of the functions $\epsilon/\bar{\epsilon}$ and $\alpha$ at the source are obtained by
\begin{align}
\frac{\epsilon}{\bar{\epsilon}}\Big|_{\mathrm{s}}
&=-\frac{\Delta \varphi_{\mathrm{p}}}{3\pi}+\frac{2m}{r}\Big|_{\mathrm{s}},
\\
\label{eq:alpNe}
\alpha(r_{\mathrm{s}}) &=r_{\mathrm{s}} 
\langle z\rangle
-\frac{m(r_{\mathrm{s}})}{2},
\end{align}
where 
$\langle z\rangle$
is the redshift of the star averaged over one orbital period, 
and we have used Eq.~\eqref{eq:zobs} to obtain Eq.~\eqref{eq:alpNe}.
We can recast the above equations to the following form:
\begin{align}
\label{eq:epNewt}
\epsilon(r_{\mathrm{s}})&=
\frac{3m^2}{2\pi r^4}\bigg|_{\mathrm{s}} \frac{
\Delta \varphi_{\mathrm{p},\mathrm{conv}}
-\Delta \varphi_{\mathrm{p}}}{
\Delta \varphi_{\mathrm{p}, \mathrm{conv}}
},
\\
\frac{\alpha-m}{m}\Big|_{\mathrm{s}}
&=\frac{3}{2} \frac{
\langle z\rangle
-\langle z\rangle_{\mathrm{conv}}
}{
\langle z\rangle_{\mathrm{conv}}},
\end{align}
where 
the symbol ``conv" stands for conventional, 
and $\Delta \varphi_{\mathrm{p}, \mathrm{conv}}$ and 
$\langle z\rangle_{\mathrm{conv}}$ are the conventional expressions for 
$\Delta \varphi_{\mathrm{p}}$ and $\langle z\rangle$, respectively, 
\begin{align}
\Delta \varphi_{\mathrm{p}, \mathrm{conv}}
&=\frac{6\pi m}{r}\Big|_{\mathrm{s}},
\\
\langle z\rangle_{\mathrm{conv}}
&=\frac{3m}{2r}\Big|_{\mathrm{s}}.
\end{align}
Thus, if we know the value of $r_{\mathrm{o}}$, we can \textit{locally} determine not only 
the deviation of $\alpha(r_{\mathrm{s}})$ from the gravitational mass $m(r_{\mathrm{s}})$ but also the energy density $\epsilon(r_{\mathrm{s}})$
at the orbital radius of the star.
The accuracy in observing 
$\langle z\rangle$
determines the sensitivity to the deviation $\alpha(r_{\mathrm{s}})$ from $m(r_{\mathrm{s}})$, while the accuracy in observing $\Delta \varphi_{\mathrm{p}}$ determines the sensitivity to $\epsilon$. If we normalize the parameters by 
typical values for S2/S0-2 near Sgr~A$^\ast$,
we obtain for the latter 
\begin{align}
\epsilon(r_{\mathrm{s}}) \approx 3.7 \times 10^{-5} M_{\odot} \mathrm{\, au}^{-3}
\left(
\frac{
\Delta \varphi_{\mathrm{p}, \mathrm{conv}}
-\Delta \varphi_{\mathrm{p}}}{
\Delta \varphi_{\mathrm{p}, \mathrm{conv}}
}\bigg/0.1
\right)\left(\frac{m}{4.0\times 10^6 M_{\odot}}\right)^2 \left(
\frac{r_{\mathrm{s}}}{120 \mathrm{\, au}}
\right)^{-4},
\end{align}
where 
$\Delta \varphi_{\mathrm{p, conv}}
\approx 0.36^{\circ}\:\! [\:\!m/(4.0\times10^6 M_\odot)\:\!](r_{\mathrm{s}}/120\mathrm{\, au})^{-1}$.
If we can implement the above 
analysis
for many stars of different $r_{\mathrm{s}}$, we can check whether such obtained functions $\alpha(r)$, $m(r)$, and $\epsilon(r)$ satisfy Eqs.~\eqref{eq:ep1} and \eqref{eq:alp1/2} and thus check the Einstein cluster solution 
as a model of the gravitational field sourced by dark matter particles surrounding the central black hole.

%%%%%
\section{Summary and discussion}
\label{sec:6}
%%%%%
We have considered 
the periapsis shift of 
geodesic bound orbits on physically reasonable static clouds in an asymptotically flat black hole spacetime. 
We ensure that the background 
spacetime constructed as the Schwarzschild black hole surrounded by a static Einstein cluster 
satisfies the four energy conditions (i.e., the weak, strong, null, and dominant energy conditions) in the entire region.
In the framework of general relativity, we have explored how 
matter distribution 
affects the prograde shift of an orbiting star (the 
periapsis shift in the same direction as the revolution) observed in vacuum black hole spacetimes.
Consequently,
we have shown that 
the precession rate $\nu$ of the nearly circular 
bound orbits
is determined by the relative magnitude of the following 
two terms of the opposite signs: 
The ratio of the gravitational radius for the 
gravitational mass 
contained within the equilibrium orbital radius to its equilibrium radius, $2m/r$, 
with a positive sign and the ratio of the local 
energy density to the averaged 
energy density within the equilibrium radius, 
$\zeta=\epsilon/\bar{\epsilon}$, 
with a negative sign. 
If the general-relativistic 
effect dominates over the 
local-density 
effect (i.e., $2m/r>\zeta$), 
then the prograde shift occurs (i.e., $0<\nu<1$),
while 
if the 
local-density 
effect dominates over the 
general-relativistic 
effect (i.e., $\zeta>2m/r$), 
then the retrograde shift occurs (i.e., $\nu<0$). 
This result means that 
the negative contribution to the 
periapsis 
shifts naturally appears as a consequence of the extended distribution of physically reasonable matters even in general relativity. 
Note that even though a retrograde shift occurs, 
it does not imply any exotic spacetime (e.g., naked-singular spacetimes or wormhole spacetimes) but simply the existence of significant local energy density on the orbit of the star. 
Furthermore, if
the prograde shift exceeds the value expected from the general-relativistic effect, it implies that the local energy density is negative, thus violating the energy conditions.

We have revealed that, 
if the distance from the observer to the star is given,
the four 
quantities
for a nearly circular 
bound orbit (i.e., the orbital shift angle, the radial oscillation period, the redshift, and the source position mapped onto the observer's sky) determine the local values of the background model functions. 
Therefore, the model functions can be extrapolated by measuring such observables with different radii. A notable advantage of focusing on nearly circular 
bound 
orbits is that the shape of the model functions can be estimated 
without 
assuming a concrete
functional form.
The discussion is much clearer for quasi-circular orbits in the post-Newtonian regime.

Furthermore, we have estimated 
the precession rate of nearly circular 
bound orbits in 
the constant density model, the isothermal sphere model, and the NFW model.
A common property of 
these models 
is that if the matter distribution is sufficiently broadened while 
the total mass 
is
fixed, the prograde 
periapsis shift occurs due to the dominant 
general-relativistic 
effect near the inner boundary of the distribution.
Conversely, the retrograde shift occurs due to the dominant 
local-density effects near the outer boundary of the distribution.
In the situation at the center of our galaxy, we find that even if the mass fraction of the matter to the black hole is only $1\%$~\cite{Heissel:2021pcw, GRAVITY:2021xju}, the 
local-density effect can compensate for the prograde shift due to the 
general-relativistic effect, which is consistent with the result in Ref.~\cite{Rubilar:2001}.
Furthermore,
if the matter distribution is localized in a narrow region while 
the total mass is fixed, the 
local-density effect tends to dominate over the 
general-relativistic 
effect, 
so that the retrograde shift occurs.
In the intermediate situation, 
there appears a variety of behaviors depending on the density distribution. 
Thus, 
we can extract information about the matter distribution from the distribution of the 
periapsis shifts.

We have also numerically 
simulated 
several bound 
orbits with 
large eccentricity
on the matter distribution. In the parameter range explored in this study, 
the shift direction remains 
unchanged
even when 
a nearly circular
bound 
orbit transits to 
a bound 
orbit with 
large eccentricity
by injecting energy. It suggests that the onset of prograde and retrograde 
periapsis 
shifts revealed by nearly circular elliptical orbits is also 
approximately valid 
even for 
bound orbits with 
large eccentricity.

This study uses a static and spherically symmetric 
black hole spacetime 
surrounded by a self-gravitating cluster of massive particles 
to consider the competing effects, the 
general-relativistic 
one and the 
local-density 
one on the 
periapsis shifts.
However, 
several future projects (e.g., Thirty Meter Telescope) are expected to achieve the accuracy to measure even the frame-dragging effect of Sgr A$^\ast$ from the 
observations of S-stars.
Therefore, 
it is an interesting future issue to
clarify the competition between the spin effects and the local-density effects for phenomena such as periapsis shifts near a rotating black hole.

Though this study 
has concerned 
the effect of local matter density distribution on the dynamics of 
stars,
the effect of matter distribution on the light rays or photon orbits is another interesting issue. We will report in a separate paper on the effect of matter distribution on phenomena such as gravitational lensing, a photon sphere, and gravitational redshifts in a black hole spacetime with static clouds.

\appendix

%%%%%
\section{Derivation of the stress-energy tensor for the Einstein cluster}
\label{sec:A}
%%%%%
We review the derivation of the stress-energy tensor for the Einstein cluster according to Einstein~\cite{Einstein:1939}.
Assume that the matter field consists of a cluster of many freely falling particles with equal mass $m_{\mathrm{p}}$. 
This means that each particle moves according to the Lagrangian~\eqref{eq:Lagra} with $m_{\mathrm{p}}$ restored. 
Then the momentum of each particle is $p_\mu=m_{\mathrm{p}} g_{\mu\nu} \dot{x}^\nu$ and must satisfy $p_\mu p^\mu=-m_{\mathrm{p}}^2$. 
This constraint equation can be separated into two equations via the separation constant 
$L_{\mathrm{p}}$
as 
\begin{align}
&\left(1-\frac{2m}{r}\right)^2 p_r^2+\left(1-\frac{2m}{r}\right)\left(
\frac{L_{\mathrm{p}}^2}{r^2}
+m_{\mathrm{p}}^2\right)=
\frac{r-2m}{r-2\alpha} \mathcal{E}_{\mathrm{p}}^2,
\\
\label{eq:Q}
&
L_{\mathrm{p}}^2
=p_\theta^2+\frac{p_\varphi^2}{\sin^2\theta},
\end{align}
where $\mathcal{E}_{\mathrm{p}}=-p_t$ is the conserved energy and
$L_{\mathrm{p}}\geq 0$
is the total angular momentum. 
Each particle 
of
the Einstein cluster moves on a circular orbit. 
Therefore, 
$L_{\mathrm{p}}$
and $\mathcal{E}_{\mathrm{p}}$ are written as functions of the orbital radius $r$, as shown in Eqs.~\eqref{eq:Lr} and \eqref{eq:Er},
\begin{align}
&
\frac{L_{\mathrm{p}}^2}{m_{\mathrm{p}}^2}
=\frac{m r^2}{r-3m},
\\
&\frac{\mathcal{E}_{\mathrm{p}}^2}{m_{\mathrm{p}}^2}=\left(1-\frac{2\alpha}{r}\right)\frac{r-2m}{r-3m}.
\end{align}

We now focus on the particles in circular motion passing through a spacetime point $x$. 
Such particles have the same value of 
$L_{\mathrm{p}}$
because they are in the same spherical layer. However, 
the pair of specific momentum components 
$p_{\theta}/L_{\mathrm{p}}$ and $p_\varphi/(L_{\mathrm{p}}\sin\theta)$
can be
distributed on the unit circle, as seen from Eq.~\eqref{eq:Q}.
Thus, we can parametrize these components by a variable $\xi$ ($0\le \xi \le 2\pi$) as
\begin{align}
\left(\frac{p_\theta(x; \xi)}{L_{\mathrm{p}}}, \frac{p_\varphi(x; \xi)}{L_{\mathrm{p}} \sin \theta }\right)
=(\cos \xi, \sin \xi).
\end{align}
We evaluate the stress-energy tensor $T_{\mu\nu}(x)$ by averaging over randomly distributed momentum in the phase space. Suppose that $T^{\mu}{}_{\nu}(x)$ takes the following form:
\begin{align}
\label{eq:Tmunuav}
T^\mu{}_{\nu}(x)=\frac{n(r)}{m_{\mathrm{p}}} \langle \:\!p^\mu p_\nu \rangle,
\end{align}
where $n(r)$ is the proper number density, and $\langle\cdot \rangle$ denotes averaging over all possible orbits passing through a point $x$. By formulating the averaging in terms of $\xi$, we can calculate Eq.~\eqref{eq:Tmunuav} for the Einstein cluster as
\begin{align}
T^\mu{}_{\nu}(x)=\frac{n}{m_{\mathrm{p}}}\frac{1}{2\pi} \int_0^{2\pi} p^\mu(x; \xi) \:\!p_\nu(x; \xi)\:\! \mathrm{d} \xi 
=\mathrm{diag} (-\epsilon, 0, \Pi, \Pi),
\end{align}
where 
\begin{align}
\epsilon&= m_{\mathrm{p}} n \left(1+
\frac{l_{\mathrm{p}}^2}{r^2}
\right),
\\
\Pi&=m_{\mathrm{p}} n 
\frac{l_{\mathrm{p}}^2}{2r^2},
\end{align}
are the energy density and the pressure uniformly applied in the tangential direction to each sphere of constant radius, respectively, where $l_{\mathrm{p}}=L_{\mathrm{p}}/m_{\mathrm{p}}$.

%%%%%
\section{Coordinates on the celestial sphere of the distant observer}
\label{sec:B}
%%%%%
We review
coordinates on the celestial sphere of the distant observer~\cite{Bardeen:1973}. 
Let $e^{(a)}$ be a natural tetrad given as
\begin{align}
e^{(0)}=\sqrt{1-\frac{2\alpha}{r}}\:\!\mathrm{d}t,
\quad 
e^{(1)}=\left(1-\frac{2m}{r}\right)^{-1/2} \mathrm{d}r,
\quad 
e^{(2)}=r\:\!\mathrm{d}\theta,
\quad 
e^{(3)}=r\sin\theta \:\!\mathrm{d}\varphi,
\end{align}
which satisfy $g_{\mu\nu}=\eta_{ab}e^{(a)}{}_\mu e^{(b)}{}_\nu$, 
where $\eta_{ab}=\mathrm{diag}(-1,1,1,1)$. Then the tetrad components of the photon momentum, $k^{(a)}=e^{(a)}{}_\mu k^\mu$, are given by
\begin{align}
k^{(0)}=\left(1-2\alpha/r\right)^{-1/2}(-k_t),
\quad
k^{(1)}=\left(1-2m/r\right)^{1/2} k_r,
\quad
k^{(2)}=k_\theta/r,
\quad
k^{(3)}=k_\varphi/(r\sin\theta).
\end{align}
These satisfy the null condition $\eta_{ab} k^{(a)} k^{(b)}=0$,
or equivalently, 
\begin{align}
\label{eq:ncondcele}
(k^{(1)}/k^{(0)})^2+(k^{(2)}/k^{(0)})^2+(k^{(3)}/k^{(0)})^2=1.
\end{align}
Using this unit sphere at infinity, 
we define angle coordinates $(\psi, \beta)$ on the celestial sphere as
\begin{align}
k^{(1)}/k^{(0)}\big|_{\mathrm{o}}=-\cos \beta,
\quad
k^{(2)}/k^{(0)}\big|_{\mathrm{o}}=\sin \beta \sin \psi,
\quad
k^{(3)}/k^{(0)}\big|_{\mathrm{o}}=-\sin\beta \cos \psi,
\end{align}
where $\beta$ and $\psi$ is the latitude and longitude, respectively, 
and the line-of-sight to the central black hole is $\beta=0$. 
We can also introduce 
another angle coordinates as 
\begin{align}
X=-\frac{k^{(3)}}{k^{(0)}}\bigg|_{\mathrm{o}},
\quad
Y=\frac{k^{(2)}}{k^{(0)}}\bigg|_{\mathrm{o}}.
\end{align}
Assume that the observer is far from the center $(r,\theta)=(r_{\mathrm{o}}, \theta_{\mathrm{o}})$, where $r_{\mathrm{o}}\gg M$. 
Then 
we can obtain the 
asymptotic forms
of $(X, Y)$ as 
\begin{align}
\label{eq:Xapp}
X&
\simeq 
-\frac{b}{r_{\mathrm{o}}\sin \theta_{\mathrm{o}}},
\\
\label{eq:Yapp}
Y&
\simeq 
\pm \frac{1}{r_{\mathrm{o}}}\sqrt{q^2-\frac{b^2}{\sin^2\theta_{\mathrm{o}}}},
\end{align}
where $b$ and $q$ are given in Eqs.~\eqref{eq:bqdef1} and \eqref{eq:bqdef2}, respectively. 
As a result, Eqs.~\eqref{eq:Xapp} and \eqref{eq:Yapp} provide the relationship between $(X,Y)$ and $(b,q)$. 

When the observer is on the axis of $\theta_{\mathrm{o}}=0$ (or $\theta_{\mathrm{o}}=\pi$), 
the coordinates $(X,Y)$ are singular. 
Furthermore, $b/\sin \theta_{\mathrm{o}}$ is indefinite for photons arriving at the axis. 
However, the latitude $\beta$ is still valid because the indefinite quantities cancel out in the expression, 
\begin{align}
\beta\simeq \sin\beta =\sqrt{(k^{(2)}/k^{(0)})^2+(k^{(3)}/k^{(0)})^2}\Big|_{\mathrm{o}}\simeq \frac{q}{r_{\mathrm{o}}}.
\end{align}
This means that the observer on the axis can relate the angle $\beta$ to $q$.

\begin{acknowledgments}
The authors are grateful to Parth Bambhaniya, Dipanjan Dey, Minxi He, 
Hideki Ishihara, Satoshi Iso, Pankaj S.~Joshi, Yasutaka Koga, Kazunori Kohri, Takahiko Matsubara, Kouji Nakamura, Ken-ichi Nakao, Keisuke Nakashi, Kenji Toma, Chul-Moon Yoo, and Hirotaka Yoshino for useful comments and helpful discussion. 
We would like to thank Andrea Maselli for their helpful comments on revisions to the manuscript. 
This work was supported by JSPS KAKENHI Grants No.~JP19K14715 and No.~JP22K03611 (TI); 
No.~JP19K03876, No.~JP19H01895, and No.~JP20H05853 (TH); No.~JP19H01900 and No.~JP19H00695 (HS, YT).
\end{acknowledgments}

\end{document}